\documentclass[sigconf]{acmart}

\setcopyright{none} 
\acmConference[Anonymous Submission to ACM CCS 2017]{ACM Conference on Computer and Communications Security}{Due 19 May 2017}{Dallas, Texas}
\acmYear{2017}

\usepackage{hypernat}
\usepackage{bmpsize}
\usepackage{textcomp}
\usepackage{epstopdf}
\usepackage{cancel}
\usepackage{soul}
\usepackage{array}
\usepackage{colortbl}
\usepackage{xspace}
\usepackage{amsmath}
\usepackage[ruled,vlined,linesnumbered]{algorithm2e}
\usepackage{pifont}
\usepackage[noend]{algpseudocode}

\newcommand{\sysname}{{Shape-GD}\xspace} 
\newcommand{\intbasis}{{\cal intuitive basis}\xspace} 

\newcommand{\ignore}[1]{}
\newcommand{\mohit}[1]{{\footnotesize\color{blue}[Mohit: #1]}}

\newcommand{\mikhail}[1]{{\footnotesize\color{red}[Mikhail: #1]}}

\newcommand{\todo}[1]{{\footnotesize\color{red}[TODO: #1]}}
\newcommand{\nonl}{\renewcommand{\nl}{\let\nl\oldnl}}
\newcommand{\codecomm}[1]{{\nonl\it\small\mdseries\textcolor{blue}{#1}}}

\makeatletter
\newcommand{\removelatexerror}{\let\@latex@error\@gobble}
\makeatother

\usepackage{url}

\begin{document}
\date{}

\title{Exploiting Latent Attack Semantics for \\ Intelligent Malware Detection}



\author{Mikhail Kazdagli, Constantine Caramanis, Sanjay Shakkottai and Mohit Tiwari}
\email{{mikhail.kazdagli, constantine}@utexas.edu}
\email{{shakkott, tiwari}@austin.utexas.edu}
\affiliation{The University of Texas at Austin\vspace{0.2in}}

\begin{abstract}

%

 

We introduce a new malware detector -- Shape-GD -- that 
aggregates per-machine detectors into a robust global detector. 
Shape-GD is based on two insights:
{\em 1. Structural:} actions such as visiting a website (waterhole attack) or
membership in a shared email thread (phishing attack) by nodes correlate well
with malware spread, and create dynamic {\em neighborhoods} of nodes that were
exposed to the same attack vector. However, neighborhoods vary unpredictably
and require aggregating an unpredictable number of local detectors' outputs into
a global alert.  {\em 2. Statistical:} feature vectors corresponding to true
and false positives of local detectors have markedly different conditional
distributions -- i.e. their {\em shapes} differ. We show that the shape of
neighborhoods can identify infected neighborhoods {\em without} having to
estimate the number of local detectors in the neighborhood.   





We evaluate Shape-GD by emulating a large community of Windows systems -- using
system call traces from a few thousand malware and benign applications -- and
simulating a waterhole attack through a popular website and a phishing attack
in a corporate email network.  In both these scenarios, we show that Shape-GD
detects malware early ($\sim$100 infected nodes in a $\sim$100K node system for
waterhole and $\sim$10 of 1000 for phishing) and robust (with $\sim$100\%
global true positive and $\sim$1\% global false positive rates). At such early
stages of infection, existing algorithms that cluster feature vectors are
ineffective (have an AUC metric of close to 0.5), and others that count the
fraction of alert-generating local detectors require (the weakly correlated)
neighborhoods' sizes to be estimated to within 1\% accuracy.


\end{abstract}



\maketitle

\section{Introduction}
\label{sec:intro}







Behavioral detectors are a crucial line of defense against malware. By
extracting features out of network packets~\cite{wenkee_lee_bothunter_2007,
paxson1999, Sommer2010, zhang2001}, system
calls~\cite{Canali2012,Christodorescu06,
 Fredrikson2010}, instruction set~\cite{Juxtapp, Christodorescu05}, and
hardware~\cite{Demme2013, Tang2014, dmitry-ensemble} level actions, behavioral
detectors train machine learning algorithms to classify program binaries and
executions as either malicious or benign.  In practice,
behavioral detectors are 
deployed extensively as 
per-machine local detectors whose alerts are analyzed by global detectors
~\cite{cisco-amp,dell-ampd,osquery,GRR,hone,hone_code}. 

\ignore{
Behavioral detectors are a widely
deployed defensive technique~\cite{cisco}
in fast-changing environments where new systems and sensors drive
previously unseen `zero-day' malware that bypass known threat models
and formal specifications -- such as abuse of new permissions in
Android~\cite{android-demystified},
row-hammer attacks on DRAMs~\cite{rowhammer_android, rowhammer_cross_vm},
accelerometers~\cite{gyrophone}
in addition to bugs in trusted codebases~\cite{master-key}.
}

However, behavioral detectors are {\em weak} -- i.e., have high false positives
and negatives. This is because a large class of malware includes benign-looking
behaviors, such as encrypting users' data, use of obfuscated code, or making
web/HTTP requests. Further, 
machine learning-based detectors have been shown to be susceptible to evasion
attacks~\cite{evade-ml,evade-pdf,practical_black_box_attacks} that either
increase false negatives or force detectors to output more false positives. As
a result, 
global detectors in enterprises with $\sim$100K local detectors have to process
millions of alerts per day~\cite{attack_graphs} which stresses heavy-weight
program analyses and human analysts who investigate the final
alerts~\cite{graphistry} -- our goal is to build a robust global detector
that amplifies weak local detectors. 


\begin{figure}[t]
\includegraphics[width=0.48\textwidth]{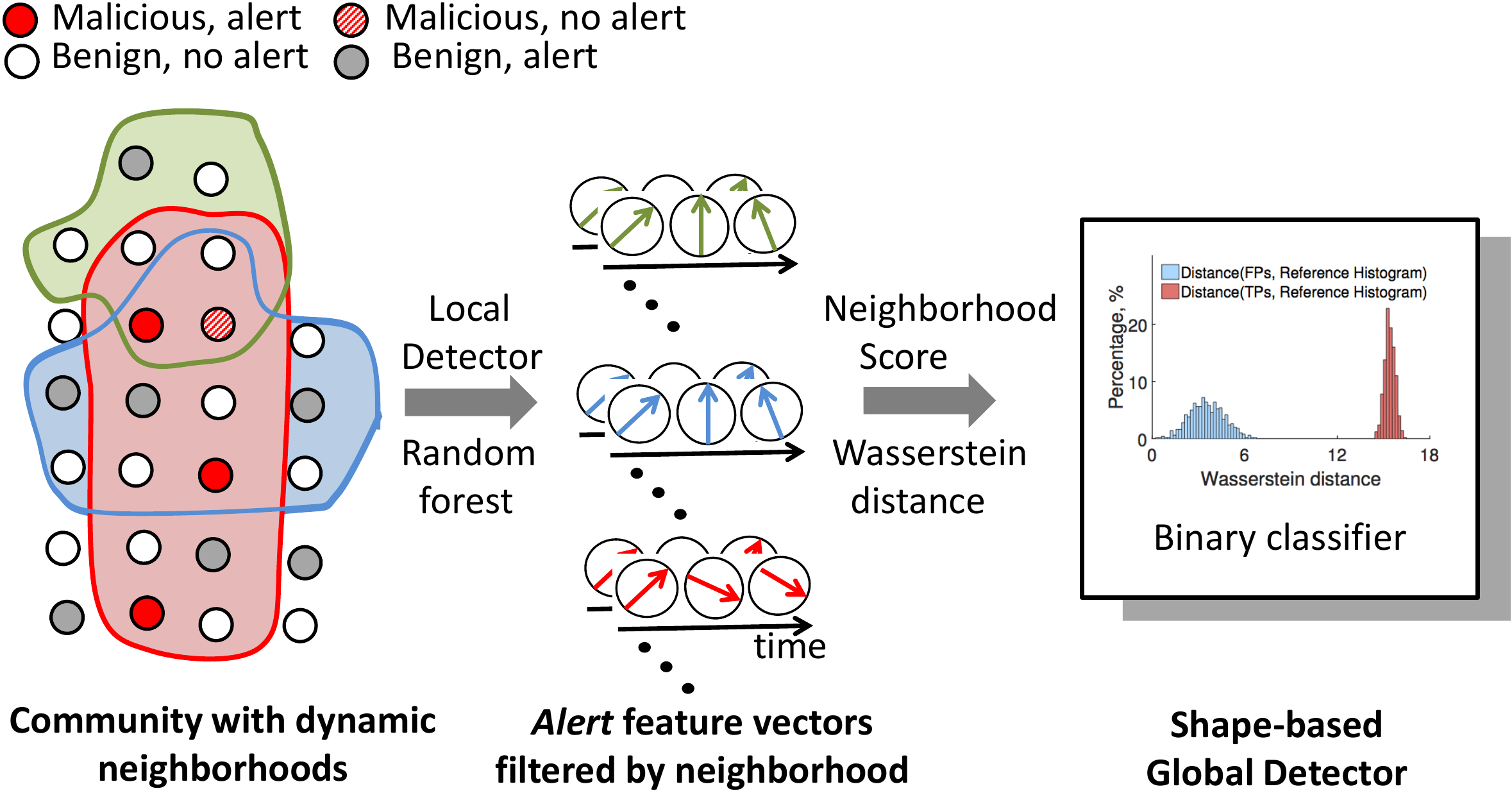}\par

\caption{\emph{(L to R) Each circle is a node that runs a local malware
detector (LD).  Our goal is to create a robust global detector (GD) from weak
LDs.  We observe that nodes naturally form {\em neighborhoods} based on
attributes relevant to attack vectors -- e.g., all client devices that visit a
website W within the last hour belong to neighborhood $NB_w$, or all users who
received an email from a mailing list M in the last hour belong to neighborhood
$NB_m$.  We propose a new GD that groups together suspicious local feature
vectors based on neighborhoods -- traditional GDs only analyze local alerts
while we re-analyze feature vectors that led to the alerts. 
Our GD then exploits a new insight -- the conditional distribution of true
positive feature vectors differs from false positive feature vectors -- to
robustly classify neighborhoods as malicious.}}

\vspace{-0.2in}
\label{fig:sys}
\end{figure}

\ignore{
Much prior research has focused on improving malware detectors deployed {\em
locally} on each machine (e.g.,
~\cite{Christodorescu:2008:MSM:1342211.1342215,kruegel-clustering,Canali2012}).
This includes engineering better features (n-grams, histograms, markov models
etc of system calls and network traffic) and composing them using ensemble
methods~\cite{dmitry-ensemble} 
}

\ignore{ Both academia and industry have realized that there exist
opportunities for correlating network-level activity and information collected
across multiple local detectors.  For example, Google is working in this
direction and has publicly released a framework for the host-level live
forensics~\cite{GRR}, however, it lacks any algorithms for automated data
analysis.  Another tool, Hone~\cite{hone,hone_code}, released by Battelle
Memorial Institute correlates network view with process-level activity on
endpoint devices, but it supports only Linux and requires deep modification of
the operating system.  }

\noindent{\bf Challenges for prior work.} Boosting weak detectors using purely
machine learning techniques is challenging.  The dominant approaches are (a)
clustering: combine feature vectors using some distance metric to identify
suspicious clusters of feature vectors~\cite{Seurat2004,beehive2013,bot_graph,bot_grep}, and (b)
counting: train local detectors (LDs) such as Random Forests to
generate local alerts, and generate a global alert if there is a significant
fraction of local alerts in the enterprise~\cite{Dash2006,BotSniffer,wenkee_lee_bothunter_2007,Shin_Infocom12_EFFORT}. 
Both approaches have limitations that force enterprises to deploy 
brittle rule-sets that explicitly correlate local detector alerts.


Clustering algorithms are well-known to be highly sensitive to noise,
especially in the high-dimensional
regime~\cite{donoho1983notion,huber2011robust,xu2013outlier}. Indeed, classical
approaches that attempt to detect or to score "outlyingness" of points (e.g.
Stahel-Donoho outlyingness, Mahalanobis distance, minimum volume ellipsoid,
minimum covariance determinant, etc) are fundamentally flawed in the
high-dimensional regime (i.e., theoretically cannot guarantee correct detection
with high probability). In practice, we see this in prior
work~\cite{beehive2013} where clustering is used primarily as a first-level
analysis to discover malicious incidents for a human analyst (i.e. requires
lower accuracy than a global detector). In Appendix~\ref{sec:eval-clustering} we
find that a clustering global detector is ineffective in early stages of
infection where our detector succeeds -- i.e., clustering yields an Area Under
Curve (AUC) metric of only $\sim$ 48\% against waterhole attack and phishing
attacks.
%
%



Count-based global detectors (Count-GD), on the other hand, suffer because they
need to know the size of local detector communities extremely accurately to
determine whether a significant fraction is raising alerts.  In practice
though, these communities of local detectors are extremely `noisy'.  For
example, consider a community of machines in an enterprise who are potentially
exposed to a so-called waterhole attack~\cite{waterhole_attack} (where a
compromised webpage spreads malicious code to machines in the enterprise).
Here, a malicious javascript-advertisement might be targeted by an ad-broker to
only a fraction of visitors to a set of webpages. Further, the specific exploit
might only succeed on a small fraction of machines that did receive the exploit
because of browser versions or patching status.  Surprisingly, our experiments
show that if Count-GD underestimates the number of nodes where the exploit ran
successfully (i.e., the community size) by just 2\%, its alerts are almost
100\% false positives (similarly, overestimating the community by 14\% leads to
almost 0\% true positives). Even small errors in estimating the number of
feature vectors in the community {\em linearly} affects the global detector's
decision thresholds.

\noindent{\bf Proposed Ideas -- Neighborhood filtering and Shape.} Our
intuition is that a weak signal indicative of malicious behavior still
separates true- and false-positive feature vectors, even though local detectors
classify both as malicious.  Our proposed system, Shape-GD, relies on two key
insights to correctly identify malicious feature vectors.

First, attack vectors into a firewalled enterprise create short-lived and
dynamic correlations across nodes -- e.g., machines that visit a specific
server (in watering hole attacks) or receive email from an address (in phishing
attacks) are more likely to be compromised than a random machine in the
community.  Since an attacker cannot target a machine inside an enterprise
directly, machines that have been exposed to a common attack vector have
correlated alerts. We call such a set of machines a {\em neighborhood}.
Neighborhoods thus concentrate the signal of malware activity
that is otherwise not visible at the overall community level and can thus
enable early detection of malware attacks.  Neighborhoods are, however,
extremely unpredictable and render cluster and count-based GDs ineffective --
hence we propose Shape-GD to aggregate local detectors' outputs.



The second insight behind Shape-GD is that the {\em distributional shape of a
set of suspicious feature vectors} can robustly separate true positive
neighborhoods from false positive neighborhoods.  Shape-GD analyzes only
those feature vectors that cause alerts by the local detectors ({\em
alert-FVs}) instead of analyzing all feature vectors. Alert-FVs thus represent
draws from one of two {\em conditional distributions} -- i.e., distribution of
malicious or benign feature vectors conditioned on being labeled as suspicious
-- which are similar but not the same.  Next, while a single suspicious feature
vector is uninformative, a set of such feature vectors can indeed be tested to
come from one of two similar-but-distinct distributions. To conduct this
hypothesis test, Shape-GD introduces a quantitative scoring function that maps a
set of feature vectors (the alert-FVs per neighborhood) into one
scalar value -- the ShapeScore of the neighborhood.  

Shape-GD composes the two insights -- i.e., filters alert-FVs along
neighborhood lines followed by computing the neighborhood's shape -- and
achieves two key properties: (i) the distribution of the alert-FVs strongly
separates malicious and benign neighborhoods (essentially, it separates the
true positive alert-FVs from false positive alert-FVs), and (ii) is robust to
noise in neighborhood size estimates (i.e., we do not need accurate
neighborhood sizes and only need a sufficient number of alert-FVs to make a
robust hypothesis test).  Specifically, \sysname detects malicious
neighborhoods with less than 1.1\% and 2\% compromised nodes per neighborhood
(in two case studies involving waterhole and phishing attacks respectively), at
a false positive and true positive rate of 1\% and 100\% respectively.
Neighborhood filtering and ShapeScore complement each other -- neighborhoods
concentrate the weak signal into a small but unpredictable set of feature
vectors while ShapeScore extracts this signal without knowing the precise
number of feature vectors. 

\noindent \textbf{Contributions.} Neighborhood filtering and shape
enable structural information about attack vectors to be captured
algorithmically. 
Our specific contributions are as follows.

\begin{itemize}

%

\item Neighborhood filtering to `reanalyze' alert feature vectors instead of
only alert time-series, and \sysname algorithms that exploit a new property --
the statistical `shape' of a neighborhood separates the ones with true
positives from those with false positives -- to classify neighborhoods
as malicious or benign without knowing their size.

\item An efficient detector that can identify malicious neighborhoods using
only 15,000 feature vectors (roughly 15 seconds of data from a 1000-node
neighborhood).  The detector comprises of random forest LDs and a Shape-GD that
computes ShapeScore as the Wasserstein distance between a set of alert-FVs'
histograms and a reference histogram built using false positive feature vectors
(created by running LDs on benign programs in uninfected machines that are used
to train the GD).

\ignore{
\item We define dynamic neighborhoods within a community of local detectors,
and introduce a new property -- the shape of a neighborhood -- that separates
true positive from false positive neighborhoods.  neighborhood filtering and
\sysname as two complementary algorithms for early and robust detection of
malware in noisy communities.

\item We implement a practical neighborhood filtering and \sysname based CIDS 
-- comprising of a random forest LD, Wasserstein distance based metric to
compute the shape of a neighborhood, and a simple threshold-based classifier to
identify malicious neighborhoods -- that can identify malicious neighborhoods
using only 15K FVs (\todo{minutes} for a 1000-node neighborhood in our
experiments). 
}



\item Phishing case study. \sysname detects
a phishing attack with 1\% false positive rate in a medium size enterprise  
network with a neighborhood of 1086 nodes when only 17.08 nodes (using temporal neighborhoods) and 4.48 nodes 
(with additional mailing-list based structural filtering) are infected.

\item Waterhole attack case study. \sysname detects a waterhole attack 
with 1\% false positive rate
when only 107.5 nodes (using temporal neighborhoods) and 139.9 (with 
additional server specific structural filtering) 
out of $\sim 550,000$ nodes are infected.


\end{itemize}

We emphasize that the LD and GD false positives (FPs) have very
different interpretations. In a phishing attack, an LD FP of 1\% in a
neighborhood of 1000 nodes means that we will get about 10 FP alerts per
second. The \sysname, on the other-hand, uses these LD FP alerts for
decision making. Thus a GD false positive occurs when it misclassifies
a neighborhood of LD alerts -- a much rarer scenario.  

Specifically, a GD FP rate of 1\% means that in our phishing attack scenario,
we will receive a global false alert about once every 100 - 300 hours.
Similarly, in the waterhole scenario a global false positive occurs every 100
sec.  Comparing the number of LDs' FPs that are reported to a GD in a Count GD
v. \sysname, temporal neighborhood filtering reduces total FPs by
$\sim$100$\times$ (phishing) and $\sim$200$\times$ (waterhole), while adding
structural filtering reduces total FPs by $\sim$1000$\times$ and
$\sim$830$\times$ respectively (Appendix~\ref{sec:discussion} for details).

Finally, as an auxiliary contribution, we present a methodology to evaluate
detectors where the LD and GD algorithms are tightly integrated. Existing
enterprise networks provide black-box LDs (such as Blue Coat, Symantec etc)
that push alert logs into `SIEM' tools (like Splunk) where GD algorithms and
visualization tools are deployed. Section~\ref{sec:exp-set} describes the
limitations of three real settings we have worked on -- a real enterprise
dataset, a university network test-bed, and the Symantec WINE dataset. None of
these allow a GD to acquire alert feature vectors from LDs.  Instead, we
incorporate a host-level malware analysis setup~\cite{kruegel-bare-metal} into real
enterprise data center~\cite{yahoo-G4} and email~\cite{enron-dataset} traces, vary the
rates of infection systematically, and thus determine the operating range of
\sysname agnostic of one specific sequence of events. This methodology offers a
more robust measure of \sysname's detection rate under adaptive malware that
can alter its infection behavior in response to \sysname's analysis.

\section{Overview of \sysname}
\label{sec:overview}

\noindent{\bf Threat model and Deployment.} We assume a standard threat model
where trusted local detectors (LDs) at each machine communicate with a trusted
global detector (GD) that receives alerts and other metadata from the local
detectors. The LDs are isolated from untrusted applications on local machines
using OS- (e.g., SELinux) and hardware mechanisms (e.g., ARM TrustZone), and
communicate with the enterprise's GD through an authenticated channel. The GD
is trained as a standard anomaly detector -- using benign data generated from
uninfected (e.g.  test/quality-assurance) machines that run LDs on benign
software, or assuming the current state of the system as benign in order to
detect future malware as anomalies.

\sysname fits deployment models that are common today.  Currently, enterprises
use SIEM tools (like HP Arcsight and Splunk) to monitor network traffic and
system/application logs, malware analysis sandboxes that scan emails for
malicious links and attachments, in addition to host-based malware detectors
(LDs) from Symantec, McAfee, Lookout, etc. We use exactly these
side-information -- from network logs (client-IP, server IP, timestamp) and
email monitoring tools -- to instantiate neighborhoods and filter LDs' alert-FVs
based on neighborhoods (Algorithm \ref{NI_algorithm}).  Upon receiving
alert-FVs, \sysname runs its malware detection algorithm (Algorithm
\ref{ShapeGD_algorithm}) for all neighborhoods the alert-FVs belong to. If a
particular neighborhood is suspicious, then \sysname will notify a downstream
analysis (deeper static/dynamic analyses or human analysts) and forward
relevant information in the incident report. 

The key difference is that \sysname needs to know the alert feature vectors from the
LDs -- black-box LDs do not currently provide these. Hence (e.g., {\tt
osquery}-based) co-designed LD-GD detectors~\cite{osquery,GRR} are the most
appropriate deployment counterpart for \sysname -- this also motivates our
experimental setup combining host-level malware analysis and web-service/email
datasets.

Operationally, the LD at each machine transforms its input signal into an alert
time series. This transformation consists of two steps: {\em (a) Generate
Feature Vectors:} convert the raw OS system calls trace into a feature vector
(FV) time series,
and {\em (b) Generate Alerts:} Determine if each FV is malicious or not using a
local detector (typically through random forests, SVM, etc.).  


\noindent{\bf Inferring neighborhoods from common attack vectors.} \sysname
operates over dynamic neighborhoods, which are updated once per neighborhood
time window (NTW).  Neighborhoods within large communities are a set of
nodes that share a statically defined {\em action attribute} within the current
time window -- this allows an analyst to create neighborhoods of nodes based on
common attack vectors.
Below are some illustrative examples of communities and neighborhoods -- we
focus our experiments on the first two examples that are responsible for a
large fraction of malware in enterprises.

\noindent {\em 1.} Waterhole attack. The community here consists of the
employees of an enterprise such as Anthem Health~\cite{anthem,
anthem-waterhole}.  In a waterhole attack, adversaries compromise a website
commonly visited by such employees as a way to infiltrate the enterprise
network and then spread within the network to a privileged machine or user.
Within this community, a neighborhood can be the set of nodes that visited the
same type of websites within the current neighborhood time window (for example,
some percentile of suspicious links rated by VirusTotal~\cite{virustotal} or
SecureRank~\cite{secure_rank}). Since these rankings themselves are fuzzy, and
the websites and their contents are dynamic, neighborhoods only indicate a
probability that the node was actually exposed to an exploit.


\noindent {\em 2.} Phishing attack over enterprise email networks.  The
community here consists of all employees within an enterprise. 
A phishing attack here would typically spread over email and use a malicious
URL to lure nodes (users) to drive-by-download
attacks~\cite{phishing-1,phishing-2} or spread through malicious attachments.
Here, a specific user's neighbors are that subset of users with whom she/he
exchanged emails with during the current neighborhood time window.

Similar correlations occur in 
physical hardware attacks -- community here consists of all
machines in a workplace that are physically proximal (e.g. machines in a
specific hospital or bank determined 
using the configuration of LAN/WiFi infrastructure, GPS information etc).
The potential attack mode here is through physical
hardware such as {\tt badUSB}. The neighborhood of a node is simply all other nodes
in the neighborhood 
that were connected to similar external hardware (e.g. a USB drive) over the
current neighborhood time window.

Attacks that target specific app-stores
(e.g., the Key-Raider attack in the Cydia app-store or the malicious
Xcode attacks due to compromised mirror sites) also propagated across
users with specific attributes (membership in a store or
downloaded Xcode from specific sites) more likely than a random user
in the network.



\subsection{Intuition behind \sysname}
\label{sec:intuition}

\begin{figure}[h]
   \centering
   \includegraphics[width=0.45\textwidth]{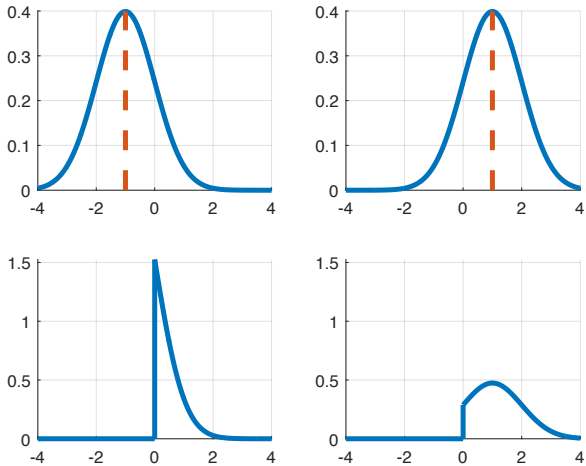}\par
   \caption{{\em (Shape of conditional distributions) The top left figure
     is the probability density function (pdf) of benign feature vectors,
     here a Gaussian with mean `-1'; and the top right figure is the
     pdf for malicious feature vectors, here a Gaussian with mean`+1'. 
     The optimal
     local detector at any machine would declare 'malware' if a sample's
     value is positive, and declare 'benign' if a sample's value is
     negative. The bottom plots shows the pdfs of the same Gaussians
     but now conditioned on the event that the sample is
     positive -- the pdfs corresponding
     to false positive and true positive feature vectors respectively 
     have different shapes.}}
\label{fig:cond-dist}
\end{figure}

The statistical shape of local detectors' false positives (FP
conditional distribution) differs from the corresponding shape for true
positives (TP conditional distribution) -- we use this property to aggregate
LDs' alert-FVs to find the shape of each neighborhood and then classify
neighborhoods based on their shapes.

%
%
%
%
%
%

The central question then is -- {\em why do true- and false-positive FVs' shapes
differ?}  To explain this and set the stage for \sysname, we consider a stylized
statistical inference example. Suppose that we have an unknown number of nodes
within a neighborhood. We want to distinguish between two extremes -- all nodes
only run benign applications (benign hypothesis), or all nodes are running
malware (malware hypothesis). We look at a single snapshot of time where each
node generates exactly one feature vector. Under the benign hypothesis, assume
that the feature vector from each node is a (scalar valued) sample from a
standard Gaussian with mean of `-1'; alternatively it is standard Gaussian with
mean of `+1' under the malware hypothesis.


{\em (a)} \underline{Noisy local detectors:} Given one sample (i.e.,
FV from one node), the {\em best} local detector is a threshold test: is
the sample's value above zero or below?  For this example, the
probability of a false positive is (about) 15\%. 

{\em (b)} \underline{Aggregating local detectors over neighborhoods:}
Suppose there are 100 nodes 
and all of them report
their value, and we are told that 90 of them are greater than 0 (i.e.,
90 of the local detectors generate alerts). In this case, the expected
number of alerts under the benign hypothesis is 15; and
85 under the malware hypothesis.  Thus, we can conclude with
overwhelming certainty ($10^{-75}$ chance of error) that 90 alerts
indicate an infected neighborhood. This corresponds to a
conventional threshold algorithm that count the number of alarms in a
neighborhood and compares with a global threshold (here this threshold
is 50).

{\em (c)} \underline{Count without knowing neighborhood size:} 
Suppose,
now, that we do not know the number of nodes (i.e., neighborhood size
is unknown), and only know that there are a total of 90 alerts. In
other words, out of the neighborhood of nodes, some 90 of them whose
samples were positive reported so. 
What can we say?  Unfortunately we
cannot say much -- if there were 100 nodes in neighborhood, then
malware is extremely likely; however, if there were 1000 total nodes,
then with 90 alerts, it is by far (exponentially) more likely that we
have no infection. Because we do not know the neighborhood size, the
global threshold cannot be computed.


{\em (d)} \underline{Robustness of Shape:} While the number of alerts
alone is uninformative, we can resolve whether the neighborhood is a
`false positive' or `true positive' by considering the actual values
of the 90 random variables corresponding to these {\em alerts}. These
values represent independent draws from a {\em conditional}
distribution -- either the distribution of a normal random variable of
mean '$-1$' {\em conditioned on taking a nonnegative value}, or the
distribution of a normal random variable of mean '$+1$' {\em
  conditioned on taking a nonnegative value} (see
Figure~\ref{fig:cond-dist}). This conditioning occurs because of the
local detector -- recall it tags a sample as an alert if and only if the
sample drawn was nonnegative (optimal LD in this example).  Thus,
irrespective of the size of the neighborhood, the global detector would
``look at the shape'' of the empirical distribution (i.e. the
distribution constructed from the received samples) of the received
samples (FVs). If this were ``closer'' to the left rather than the
right plot in Figure~\ref{fig:cond-dist}, it would declare
``uninfected''; otherwise declare ``infected''.



%


\subsection{From Intuition to Algorithm Design} 

Interestingly, we show that the intuition behind this simple example scales to
real malware detectors that use high-dimensional feature vectors. However, 
to use this insight in practice, we need to address two issues: {\em (i)} while
the two figures in Figure~\ref{fig:cond-dist} are visually distinct, an
algorithmic approach requires a quantitative score function to separate between
the (vector-valued) conditional distributions generated from feature vector
samples; and {\em (ii)} the global detector receives only finitely many
samples; thus, we can construct (at best) only a noisy estimate of the
conditional distribution.  We describe Shape-GD's details in
Section~\ref{sec:model} but present the key intuition here.


We introduce {\bf ShapeScore} -- a score function based on the Wasserstein
distance \cite{wasser-wiki} to resolve between conditional distributions. We
choose Wasserstein distance because it has well-known robustness properties to
finite-sample binning \cite{vallender1974calculation,benbre00}, was more
discriminative than L1/L2 distances in our experiments, and yet is efficient to
compute for vectors. 

Specifically, given a collection of feature vector samples, we construct an
empirical (vector) histogram of the FV samples, and determine the Wasserstein
distance of this histogram with respect to a {\em reference histogram}.  This
reference histogram is constructed from the feature vectors corresponding to
the {\em false positives} of the local detectors. In other words, this
reference histogram captures the statistical shape of the ``failures'' of the
LDs -- i.e., those FVs that the LD classifies as malicious even though they
arise from benign applications. Computing the reference histogram does {\em
not} require analysts to manually label alert-FVs as false positives -- these
can be generated by running the LDs on benign software in uninfected machines
(e.g., test or quality-assurance machines, by recording and replaying real user
traces on benign applications on training servers, etc). Alternatively,
analysts can use applications deployed currently and recompute the reference
histogram periodically -- this is similar to anomaly detectors where the goal
is to label anomalous behaviors as (potentially) malicious.


If we had the idealized scenario of infinite number of feature vector samples,
the ShapeScore would be uniquely and deterministically known. In practice
however, we have only a limited number of feature vector samples; thus
ShapeScore is noisy.  Our experiments (Figure~\ref{fig:windows-hist}) test its
robustness with Windows benign and malicious applications (Section~\ref{sec:results}), 
where the ShapeScore is computed from
neighborhood sizes of 15,000 FVs (about 15 seconds of data from 1000 nodes).
The key result is the strong statistical separation between the
ShapeScores for the TP and FP feature vectors respectively, thus lending
credence to our approach.  Importantly, both these ideas do not depend on
knowing the neighborhood size; thus they provide a new lens to study malware at
a global level.


\section{Related Work}
\label{sec:related-work}

\subsection{Behavioral analysis}

Behavioral analysis refers to statistical methods that monitor signals from
program execution, extract features and build models from these signals, and
then use these models to classify processes as malicious.  Importantly, as we
discuss in this section, all known behavioral detectors have a high false
positive and negative rate (especially when zero-day and mimicry
attacks are factored in).


System-calls and middleware API calls have been studied extensively 
as a signal 
for behavioral detectors~\cite{Forrest1996,
Wagner2002,
kruegel2007,
Christodorescu:2008:MSM:1342211.1342215,
Fredrikson2010,
Canali2012, robertson_anomaly}.
Network intrusion 
detection systems~\cite{paxson1999}
analyze network traffic to detect known malicious
or anomalous behaviors.
More recently, behavioral detectors use signals such as power
consumption\cite{clark2013}, CPU
utilization, memory footprint, and hardware
performance counters\cite{Demme2013,Tang2014}. 


Detectors then extract {\em features} from these raw signals.  For example, an
n-gram is a contiguous sequence of n items that 
captures total order relations~\cite{BotSniffer,Canali2012}, n-tuples are ordered
events that do not require contiguity, and bags are simply histograms.
These can be combined to create bags of tuples, tuples of bags, and tuples of
n-grams~\cite{Canali2012,Forrest1996}
often using principal component analysis to reduce dimensions.
Further, system calls with their arguments form a dependency graph structure 
that can be compared to sub-graphs that represent malicious
behaviors~\cite{Christodorescu:2008:MSM:1342211.1342215,Kolbitsch:2009:EEM:1855768.1855790,kruegel2007}.

Finally, detectors train models to classify executions into malware/benignware
using supervised (signature-based) or unsupervised (anomaly-based) learning.
These models range from distance metrics, histogram
comparison, hidden markov
models (HMM), and neural networks (artificial neural
networks, fuzzy neural networks, etc.), to 
more common classifiers such as
kNN, one-class SVMs,
decision trees, and ensembles thereof. 


Such machine learning models, however, result in high false positives and
negatives.  
Anomaly detectors can be circumvented by mimicry attacks where
malware mimics system-calls of benign applications~\cite{Wagner2002} or
hides within the diversity of benign network traffic\cite{Sommer2010}.
Sommer et al.~\cite{Sommer2010} additionally highlights 
several problems that can arise due
to overfitting a model to a non-representative training set, suggesting signature-based
detectors as the primary choice for real deployments. Unfortunately, 
signature-based detectors cannot detect new (zero-day) attacks.
On Android, both system
calls~\cite{Burguera2011} and hardware-counter based
detectors~\cite{Demme2013} 
yield $\sim$20\% false positives and $\sim$80\% true positives.

Finally, with their
ability to extract highly effective features, deep nets {\em may} provide a
new way forward for creating novel behavioral detectors. At the global
level, however, what is needed is a data-light approach for global
detection by composing local detectors, tailored to be agile enough to
do global detection in a fast-changing (non-stationary) environment.

\subsection{Collaborative Intrusion Detection Systems (CIDS)}
\label{sec:cids}


Collaborative intrusion detection systems (CIDS) provide an architecture where
LDs' alerts are aggregated by a {\em global detector} (GD).
GDs can use either signature-based or 
anomaly-based\cite{zhang2001,vlachos2004}, or even a combination
of the two~\cite{krugel2001} to generate global alerts.
Additionally, the CIDS architecture can be centralized, hierarchical,
or distributed (using a peer-to-peer overlay network)~\cite{zhang2001}.

In all cases, existing GDs use some variant of either clustering or count-based
algorithms to aggregate LDs' alerts. Count-based GD raises an alert once the
number of alerts exceeds a threshold within a space-time window, while
clustering-based GD may apply some heuristics to control the number of alerts
~\cite{Dash2006,Seurat2004,BotSniffer,wenkee_lee_bothunter_2007,Shin_Infocom12_EFFORT}. 
In HIDE~\cite{zhang2001}, 
the global detector at each hierarchical-tier is a neural network trained on
network traffic information.  Worminator\cite{Locasto2005} additionally uses
bloom filters to compact LDs' outputs and schedules LDs to form groups in order
to spread alert information quickly through a distributed system.  All count-
and clustering-based algorithms are fragile when the noise is high (in the
early stages of an infection) and when the network size is uncertain.
In contrast, our neighborhood filtering and shape-based GD is robust against
such uncertainty.


Note that distributed CIDSs are vulnerable to probe-response attacks, where the
attacker probes the network to find the location and defensive capabilities of
an LD~\cite{Shmatikov2007,Bethencourt2005,Shinoda2005}. These attacks are
orthogonal to our setting since we do not have fixed LDs (i.e. all nodes are
LDs).

\section{Shape GD Algorithm}
\label{sec:model}

Our algorithm consists of feature extraction, local detector (LD), and the
global detector (GD). Our key innovations are in the GD, however, 
we also discuss feature extraction and LD design for completeness.
\ignore{Our key innovations are in the GD and we discuss feature
extraction and LDs further in}

\begin{figure}
\removelatexerror
\begin{algorithm}[H]
 \small
 \DontPrintSemicolon
  \SetKwInOut{Input}{Input}
 \SetKwInOut{Output}{Output}
 \Input{Sequence of executed system calls}
 \Output{Alert-FVs}
 Let $id$ be LD's identifier\;
 \BlankLine
 
 \While{True}{
   
   $syscall$-$hist$ $\leftarrow$ $r$-sec histogram of system calls\;
     
   $syscall$-$hist_{PCA}$ $\leftarrow$ project $syscall$-$hist$ on $L$-dim PCA basis\;
     
   $label$ $\leftarrow$ BinaryClassifier($syscall$-$hist_{PCA}$);
  \BlankLine

   \If{$label = malicious$}{
     $alert$-$FV$ $\leftarrow$ $syscall$-$hist_{PCA}$\;
     send $<alert$-$FV$, $id>$ to \sysname\;
   }
 }
 \caption{Local Detector}
 \label{LD_algorithm}
\end{algorithm}
 \vspace{-0.2in}
\end{figure}

\noindent {\bf Feature Extraction algorithm.} This algorithm transforms the
continuously evolving 390-dimensional time-series of Windows system calls into
a discrete-time sequence of feature-vectors (FVs). 
This is accomplished by chunking the continuous time series into $r-$second
intervals, and representing the system call trace over each interval as a
single $L-$dimensional vector (Algorithm \ref{LD_algorithm}, lines 3--4). $L$
is typically a low dimension, reduced down from 390 using PCA analysis, to (in
our experiments) $L=10$ and $r = 1$ second.


\noindent {\bf Local Detector (LD) Algorithm.} The LD algorithm (Algorithm
\ref{LD_algorithm}) leverages the current state-of-the-art techniques in
automated malware detection to generate a sequence of {\em alerts} from the FV
sequence. Specifically, using both its internal state and the {\em current} FV,
the LD algorithm generates an alert if it thinks that this FV corresponds to
malware, and produces no alert if it thinks that the current FV is benign
(lines 5--8). Henceforth, we define an {\bf alert-FV} to be an FV that
generates an LD alert (either true or false positive). In our experiments each
LD employs Random Forest as a binary classifier for malware detection because
Random Forest achieves the best performance on the training data set 
among the classifiers from a prior survey~\cite{Canali2012}  and
have been shown to be robust to adversarial inputs -- we pick an operating
point of 92.4\% true positives at a false positive rate of 6\% from the LD's
ROC curve (which is similar to detection rates in prior work~\cite{Canali2012}).  
We have described our experiments with picking the best LD from a
prior survey~\cite{Canali2012} in Appendix~\ref{sec:LDs} and Figure
\ref{fig:windows-LD-roc}. 

\noindent {\bf Neighborhood Instances from Attack-Templates.} Each neighborhood
time window (NTW), Shape GD generates neighborhood instances
(Algorithm~\ref{NI_algorithm}) based on statically defined attack vectors --
each attack vector is a ``Template'' to generate neighborhoods with.
Algorithm~\ref{NI_algorithm} shows
how the concept of neighborhood unifies operationally distinct attacks like
waterhole and phishing. 


The template for detecting a waterhole attack forms a neighborhood out of
client nodes that access a server or a group of servers within a neighborhood
time window. The other template, which is used for detection of a
phishing attack, includes in a neighborhood email recipient machines belonging
to a set of mailing lists.  The two templates are shown in lines 3--10. 
\begin{figure}
\removelatexerror
\begin{algorithm}[H] 
 \small
 \DontPrintSemicolon
 \newcommand\mycommfont[1]{\footnotesize\ttfamily\textcolor{blue}{#1}}
 \SetCommentSty{mycommfont}
 \SetKwInOut{Input}{Input}
 \SetKwInOut{Output}{Output}
 \Input{Template-type, NTW}
 \Output{Set of active neighborhoods NBDs}
 \BlankLine
 \codecomm{[\textit{time, time+NTW}] \textit{defines the current time window}}\;
 \textit{time} $\leftarrow$ \textit{current time}\;
 \BlankLine
 
\While{True}{
 \uIf{Template-type = $waterhole\ attack$}{
	$V$ := client machines\codecomm{*}\;
	$S$ := accessed servers\codecomm{*}\;
 	predicate(A:Client, B:Servers) := $A$ accesses $B$\;
 	\BlankLine
 }
 \uElseIf{Template-type = $phishing\ attack$}{
 	$V$ := email recipient machines\codecomm{*}\;
 	$S$ := mailing lists\codecomm{*}\;
 	predicate(A:Recipient, B:Mailing list) := $A$ $\subseteq$ $B$\;
 }
 \BlankLine

 \codecomm{partitioning a set into non-disjoint sets to incorporate}\;
 \vspace{-.03in}
 \codecomm{structural filtering}\;
 ${P_{1}, P_{2}, ..., P_{N}}$ $\leftarrow$ partition-set($S$), where $S$ = $\bigcup\limits_{i=1}^{N}    P_{i}$\;
 
 \codecomm{form neighborhoods $NB_i$ using partitions $P_i$}\;
 $NB_i$ $\leftarrow$ $\{V |$ predicate($V$, $P_i$)$\}$\;
 
 \codecomm{set expiration time for a neighborhood $NB_i$}\;
 $NB_i$.expiration-time $\leftarrow$ \textit{t+NTW}\;
 
 \codecomm{add all neighborhoods to the set $NBDs$}\;
 $NBDs$ $\leftarrow$ $\{NB_i$ $|$ $\forall i$ in $[1,N]\}$\;
 
 \codecomm{advance time by NTW sec}\;
 \textit{time} $\leftarrow$ \textit{time+NTW}\;
 }
 \BlankLine
 \codecomm{*active within the time window [\textit{time, time+NTW}]}\;
 \caption{Neighborhoods from Attack-Vectors}
 \label{NI_algorithm}
\end{algorithm}
 \vspace{-0.2in}
\end{figure}


For simplicity we present a batch version of the neighborhood instantiation
algorithm (Algorithm \ref{NI_algorithm}) which advances time by NTW and creates
new neighborhoods for each NTW.
In contrast, the online Shape GD version updates already existing
neighborhoods while monitoring client--server interactions in real-time -- we
demonstrate the online Shape GD algorithm to detect waterhole attacks 
and the batch version against phishing attacks 
in our evaluation.  



The neighborhood instantiation algorithm accepts a template type as input, i.e.
either a template for detecting a waterhole attack or a phishing attack, and
duration of an NTW. 
The algorithm runs once per NTW -- starting by defining the sets $V$ and $S$ that will
be used to form neighborhoods.  For a waterhole attack, the set
$V$ includes all client machines accessing a set of servers and $S$ is a set of
the accessed servers. To instantiate neighborhoods for a phishing attack, 
$V$ is a set of all email recipient machines and $S$ is a set of mailing lists.
In both cases the algorithm considers only the entities that are active within
a current NTW window.

Each attack requires a predicate that determines relation between the elements
of the sets $V$ and $S$. For a waterhole attack such a predicate is true if a
client \textit{accesses} one of the servers (line 6). In the case of a phishing
template, the predicate is evaluated to true if a recipient \textit{belongs} to
a particular mailing list (line 10).

The neighborhood instantiation algorithm proceeds with partitioning the set $S$
into one or more disjoint subsets $P_i$ (line 11). This is to incorporate
`structural filtering' into the algorithm, allowing an analyst to create
neighborhoods based on subsets of servers (instead of all servers in case of
waterhole) or divide all mailing lists into subsets of mailing lists (in the
phishing).  Structural filtering boosts detection under certain conditions (see
Section \ref{sec:time-struct}). 


The neighborhood instantiation algorithm builds a neighborhood for each partition
$P_i$ using a corresponding predicate (line 12).  After forming a
neighborhood, the algorithm sets its expiration time (line 13), which is the
end of the current NTW window. All the neighborhoods in the set $NBDs$
are discarded at the end of the current NTW window.
Finally, the algorithm adds the just formed neighborhoods to $NBDs$ (line
14) and advances time by one NTW (line 15). 

The template-based neighborhood instantiation algorithm (Algorithms
\ref{NI_algorithm}) 
shares the $NBDs$ data structure with the 
Algorithm \ref{ShapeGD_algorithm}
that uses neighborhoods' shapes to detect malware.

%


\noindent {\bf Malware Detection in a Neighborhood.} 
Algorithm~\ref{ShapeGD_algorithm} detects malware {\em per neighborhood}
instead of individual nodes. 
%
%
The input to the algorithm is a set of alert-FVs from each neighborhood and its
output is a global alert for the neighborhood.
We now describe how the algorithm distinguishes 
between the {\em conditional distributions} of alert-FVs from true-positive and
false positive neighborhoods.

The key algorithmic idea is to first extract neighborhood-level features --
i.e., to map all alert-FVs within a neighborhood to a {\em single}
vector-histogram which robustly captures the neighborhood's statistical
properties.  Then, Shape GD compares this vector-histogram to a reference
vector-histogram (built offline during training) to yield the neighborhood's
ShapeScore.  The reference vector-histogram is constructed from a set of false
positive alert-FVs -- thus, it captures the statistical shape of
misclassifications (FPs) by the LDs but at a neighborhood scale. Finally,
Shape-GD trains a classifier to detect anomalous ShapeScores as malware.  This
is a key step in Shape-GD -- i.e., mapping alert-FVs from a neighborhood into a
{\em single} vector-histogram and then into a discriminative yet robust
ShapeScore lets us analyze the joint properties of all alert-FVs generated
within a neighborhood without requiring the FVs to be clustered or alerts to be
counted.  We describe these steps in further detail.

\begin{figure}
\removelatexerror
\begin{algorithm}[H]
\small
 \DontPrintSemicolon
 \SetKwInOut{Input}{Input}
 \SetKwInOut{Output}{Output}
 \Input{$L$-dim projections of alert-FVs}
 \Output{Malicious neighborhoods}
 Let $NBDs$ be a set of neighborhoods\;
 \BlankLine

 \For{each NB in NBDs}{
   \codecomm{aggregate L-dim projections of alert-FVs on per neighborhood basis}\;
  $B$ $\leftarrow$ $\{alert-FVs$  $|$ node $id$ $\subseteq$ $NB\}$\;


   \codecomm{build an $(L, b)$-dim. vector-histogram}\;
   $H_B \leftarrow$ bin \& normalize $B$ along each dimension\;

   \codecomm{compute a neighborhood score -- ShapeScore}\;
   $ShapeScore$ $\leftarrow$ Wasserstein Dist.($H_B$, $H_{{\rm ref}}$)\;

   \codecomm{perform hypothesis testing}\;
   \If{$ShapeScore > \gamma$}{
     label $NB$ as \textit{malicious}\;
   }
 }
 \caption{Malware Detection in a Neighborhood}
 \label{ShapeGD_algorithm}
\end{algorithm}
 \vspace{-0.2in}
\end{figure}

\noindent {\bf Generating histograms from alert-FVs.} The algorithm aggregates
$L$-dimensional projections of alert-FVs on per neighborhood basis into a set
$B$ (Algorithm \ref{ShapeGD_algorithm}, line 3).
After that, Shape GD converts low dimensional representation of alert-FVs, the
set $B$, into a single $(L, b)$-dimensional vector-histogram denoted by $H_B$
(line 4).  The conversion is performed by binning $L$-dimensional vectors
within the $B$ set along each dimension.  In each of the L-dimensions, the
scalar-histogram of the corresponding component of the vectors is binned and
normalized.  Effectively, a vector-histogram is a matrix $L$x$b$, where $L$ is
the dimensionality of alert-FVs and $b$ is the number of bins per dimension. 

We use standard methods to determine the size and number of bins and note that
the choice of Wasserstein distance in the next step makes Shape GD robust
against variations due to binning.  In particular, we tried square-root choice,
Rice rule, and Doane's formula~\cite{binning} to estimate the number of bins,
and we found that 20--100 bins yielded separable histograms (as in Figure
\ref{fig:windows-hist}) for the Windows dataset and fixed it at 50 for our
experiments.

%

\noindent {\bf ShapeScore.} 
We get the ShapeScore by comparing this histogram, $H_B$, to a {\em reference
histogram}, $H_{{\rm ref}}$,  which is generated during the training phase
using only the false positive FVs of the LDs.  We run LDs on the system-call
traces generated by benign apps -- the FVs corresponding to the alerts from the
LD (i.e., the false positives) are then used to construct the reference
histogram $H_{{\rm ref}}.$ ShapeScore is thus the distance of a neighborhood
from a benign reference histogram -- a high score indicates potential malware.

To collect known benign traces, a straightforward approach is to
use test inputs on benign apps or use record-and-replay tools to re-run
real user inputs in a malware-free system.  Or, like any anomaly detector, an
enterprise can train \sysname using applications deployed currently and recompute
$H_{ref}$ periodically. 


The ShapeScore of the accumulated set of FVs, $B$,
is given by the sum of the coordinate-wise Wasserstein distances
\cite{vallender1974calculation} (Algorithm \ref{ShapeGD_algorithm}, line 5) between 
$$H_B =
(H_{B}(1) \ H_{B}(2) \ \ldots \ H_{B}(L))$$ 
and 
$$H_{{\rm ref}} = (H_{{\rm ref}}(1) \ H_{{\rm ref}}(2) \ \ldots \
H_{{\rm ref}}(L)).$$ 
In other words,
$$
{\rm ShapeScore} = \sum_{l=1}^L d_W(H_{B}(l),H_{{\rm ref}}(l)),
$$
where for two scalar distributions $p, q,$ the Wasserstein distance
\cite{vallender1974calculation,wasser-wiki} is given by
$$
d_W(p, q) = \sum_{i=1}^b \left |\sum_{j=1}^i \left(p(j) - q(j) \right) \right |.
$$

This Wasserstein distance serves as an efficiently computable one dimensional
projection, that gives us a discriminatively powerful metric of distance
\cite{vallender1974calculation,benbre00}. Because the Wasserstein distance
computes a metric between distributions -- for us, histograms normalized to
have total area equal to 1 -- it is invariant to the number of samples that
make up each histogram. Thus, unlike count-based algorithms, {\em it is robust
to estimation errors in community size}. Figure~\ref{fig:windows-hist} verifies
this intuition, and shows that true positives and false positive feature
vectors separate well when viewed through the ShapeScore. 

Finally, to determine whether a neighborhood has malware present we perform
hypothesis testing. If ShapeScore is greater than a threshold $\gamma$, we
declare a global alert, i.e., the algorithm predicts that there is malware in
the neighborhood (lines 6--7).  The robustness threshold $\gamma$ is computed
via standard confidence interval or cross-validation methods with multiple sets
of false-positive FVs (see Section \ref{sec:power-shape}).

\noindent \textbf{Computing Shape GD's parameters.} 
Here we elaborate on the steps that should be taken in a real world environment
to choose parameters. The steps discussed here are generic and are applicable
to other attacks beyond waterhole and phishing -- the following results section
quantifies each of these steps.

First, an analyst should start with designing an appropriate algorithm
to run on local detectors (LDs) (Appendix \ref{sec:LDs}). To achieve
this, an analyst needs to compare the performance of multiple feature
extraction (FE) algorithms combined with a diverse set of machine
learning classifiers. One way to choose the best pair of a FE
algorithm and a classifier is to build ROC curves for each pair, and
select the pair that meets the desired detection rate to 
computation/training effort for the LD.



Second, the analyst needs to determine 
whether even a purely malicious neighborhood can be separated from
benign ones, and 
the minimum number of
FVs per neighborhood to do so (Sections~\ref{sec:power-shape} and
Appendix~\ref{sec:nbd-size}). This
number depends on the false positive rate of LDs ( e.g., in our
experiments, we determined that a neighborhood should generate at
least 15K FVs, see Figure \ref{fig:windows-hist-stability}).

Third, we need to choose an NTW based on the false positive rate (FP)
and the desired time-to-detection (Section~\ref{sec:time-nf}. 
A small NTW means more frequent
transfers of FVs from LD to GD, whereas a long NTW means that more
nodes can get compromised before the GD makes a decision and/or FPs
can drown out TPs. Similarly, structural filtering can improve 
detection rate if the true positive alert-FVs are not deluged
by the rate of false-positive alert-FVs -- Section~\ref{sec:time-struct}
quantifies how this trade-off differs for waterhole and phishing.

\section{Experimental Setup}
\label{sec:exp-set}

\subsection{Case for a New Methodology}
\label{sec:exp_justification}

\sysname experiments require datasets where the global detector can acquire
alert-FVs from local detectors, similar to {\tt osquery}-based systems where
the LD and GD are co-designed. We describe our experience with three existing
methodologies and datasets -- none of them allow alert-FVs to be acquired,
provide complete ground truth infection information, or allow the infection
rate to be varied. This motivates the methodology we use to systematically
evaluate \sysname. 

\noindent{\bf Analysis of existing datasets.}
Prior work has used enterprise logs~\cite{mining_log_data_alina_oprea} that are
unavailable publicly.  We have acquired similar security logs from a Fortune 500
company with a 200K machine network -- the logs average 250M entries per day
over a 2 year period, arise from  20 closed-source endpoint local detectors
such as Symantec, McAfee, Blue Coat, etc, have almost 500 sparsely populated
dimensions per log entry, and about 75\% of the log entries lack important
identity and event-timestamp information and are delayed by up to 60 days.
Commercial (black-box) LDs do not expose feature vectors for external analysis.

We have also acquired network {\tt pcap} traces from our university network,
emulating prior work~\cite{network_intrusion_detection} in network-only global detectors.
University security groups (like ours) are only allowed to collect
network-layer {\tt pcap} information for a rolling 2-week period and cannot
instrument host machines (that are owned by students and visitors) -- i.e., our
4TB/day dataset from 150K machines is unsuitable to evaluate \sysname because
it doesn't have LDs. Extending this dataset with a weak LD -- the ability to
inspect executables in a sandbox downloaded by hosts (e.g., as pursued by
Lastline~\cite{lastline}) -- would be an appropriate experimental setup but
sensitive data issues make such datasets hard to get. Hence, we model this
extended setup in our methodology. 

We have analyzed Symantec's WINE dataset~\cite{downloader_graphs} and found it inadequate to
evaluate \sysname even after layering VirusTotal information on it.
Specifically, the WINE dataset includes downloader
graphs~\cite{downloader_graphs} -- the nodes are executables and the edges
represent whether the source downloaded the destination executable -- and
represent downloader trojans (`droppers') in malware distribution
networks~\cite{malware_distribution_net} that download payloads to steal
information, encrypt the disk, etc on to the host.  This dataset covers 5 years
of data with 25M files (specifically, hashes that represent files) on over 1M
machines -- however, only 1.5M of the 25M hashes have reports on VirusTotal.
Hence, one cannot reconstruct (alert) feature vectors for the hashes stored in
the downloader graph.

\sysname re-analyzes (local) alert feature vectors in the global detector --
filtering alert-FVs into neighborhoods and then computing the neighborhoods'
shapes.  
Hence we model {\tt osquery}-like
deployments as used in enterprises like Facebook and Google where the LDs and
GDs are co-designed and GDs can acquire alert-FVs. 

\noindent{\bf Simulating malware propagation in a network.} Methodologies that
use existing datasets with malware propagation (like the ones above) have an
inherent weakness. Such datasets have one sequence of malware propagation
events ``hardwired'' into the dataset and do not allow us to analyze how a
detection mechanism reacts to variations of malware propagation dynamics,
especially when malware can adapt these dynamics.  Instead, we propose to vary
the rate of infection (which changes the neighborhood formation) and determine
\sysname's detection performance across different infection rates.


Further, none of the above datasets provide ground truth information about the
true extent of infections, 
incentivizing a design that minimizes false positives at the expense of false
negatives. In a controlled setting where host-level malware and benignware traces 
are overlaid onto a trace of web-service/email communication, we can maintain
ground truth information and determine false positives and negatives precisely.

To this end, we use a malware and benignware dataset from a recent related
work~\cite{kruegel-bare-metal}, 
train an LD with histogram-based feature vectors and a Random Forest detector
based on a recent survey on host-level malware detection~\cite{Canali2012},
and overlay these host-level results on web-service (network) and email traces
using two standard (and publicly available) datasets from Yahoo data centers
and Enron respectively. We describe this methodology in detail next.

\subsection{Benign and Malware Applications}
\label{sec:apps}


We collect data from thousands of benign applications and malware samples.  To
avoid tracing program executions where malware may not have executed any stage
of its exploit or payload correctly, we set a threshold of 100 system calls per
execution to be considered a success.  Our experiments successfully run 1,311
malware samples from 193 malware families collected in July
2013~\cite{kruegel-bare-metal}, and 2,364 more recent samples from 13 popular
malware families collected in 2015 \cite{kaspersky-windows-2015}, to compare
against traces from 1,889 benign applications.




We record time stamped sequences of executed system calls using Intel's Pin
dynamic binary instrumentation tool.  Each Amazon AWS virtual machine instance
runs Windows Server 2008 R2 Base on the default T2 micro instances with 1GB
RAM, 1 vCPU, and 50GB local storage.  The VMs are populated with user data
commonly found on a real host including PDFs, Word documents, photos, Firefox
browser history, Thunderbird calendar entries and contacts, and social network
credentials.  To avoid interference between malware samples, we execute each
sample in a fresh install of the reference VM.
As malware may try to propagate over the local network, we set up a sub-net of
VMs accessible from the VM that runs the malware sample.
In this sub-net, we left open common ports (HTTP, HTTPS, SMTP, DNS, Telnet, and
IRC) used by malware to execute its payload. We run each benign and malware
program 10 times for 5 minutes per run 
for a total of almost 53,000 hours total compute time on Amazon AWS.


Overall, benignware and malware were active for 141,670 sec and 283,270 seconds
respectively, executing an average of 11,900 and 13,500 system calls per second
respectively.  Using 1 second time window (Section \ref{sec:model})
and sliding the time windows by 1ms, we extract histograms of system calls
within each time window as the ML feature, and finally pick 1.5M benign and 1M
malicious FVs from this dataset for the experiments that follow. Importantly,
we do not constrain the samples on neighboring machines to belong to the same
families -- as described above, 
malware today predominantly spreads through malware distribution networks
where a downloader trojan (`dropper') can distribute arbitrary and unrelated 
payloads on hosts. We want to test \sysname in the extreme case
where malicious FVs can be assigned from any malware execution to any machine.


%


\subsection{Modeling Waterhole and Phishing Attacks}
\label{sec:datasets}

\noindent \textbf{Waterhole attack.} To model a waterhole attack, we use
Yahoo's ``G4: Network Flows Data"~\cite{yahoo-G4} dataset, which contains
communication data between end-users and Yahoo servers.  The 41.4 GB (in
compressed form) of data were collected on April 29-30, 2008.  Each netflow
record includes a timestamp, source/destination IP address, source/destination
port, protocol, number of packets and the number of bytes transferred from the
source to the destination. 
We model the setting where heuristics such as SecureRank~\cite{secure_rank} are
applied to identify suspicious servers and 
we assume that \sysname monitors the top (here, 50) suspicious servers based on
SecureRank's scores.  Specifically, we use 5 hours of network traffic (208
million records) captured on April 29, 2008 between 8 am and 1 pm at the border
routers connecting Dallas Yahoo data center (DAX) to the Internet. 
The selected 50 DAX servers communicate with 3,181,127 client machines over
14,249,931 requests.

We assume that an attacker compromises one of the most frequently accessed DAX
server -- 118.242.107.76, which processes $\sim 752,000$ requests within 5-hour
time window ($\sim 43.7$ requests per second).  In our simulation it gets
compromised at random instant between 8am and 10.30am.  Hence, Shape GD can use
the remaining 2.5 hours to detect the attack (our results show that less than a
hundred seconds suffice).  Following infection, we simulate this `waterhole'
server compromising client machines over time with an infection probability
parameter -- this helps us determine the time to detection at different rates
of infection.  The benign and compromised machines then select corresponding
type of execution trace (i.e., a sequence of FVs generated in
Section~\ref{sec:apps}) and input these to their LDs.


\noindent \textbf{Phishing attack.} We simulate a phishing attack in a medium size
corporate network of 1086 nodes that exchange emails with others in the
network.  To model email communication, we pick 50 email threads with 100
recipients each from the publicly available Enron email dataset
\cite{enron-dataset} (the union of all email threads' recipients is 1086). 

We start the simulation with these 50 emails being sent into the 1086-node
neighborhood, and seed only {\em one} email out of 50 as malicious.
We then model the infection speading at different rates as this malicious email
is opened by its (up to 100) recipients at some time into the simulation and is
compromised with some likelihood when the user `clicks' on the URL in the
email.  Our goal is to measure the number of compromised nodes before Shape GD
declares an infection in this neighborhood. All nodes that open and `click' the
link in the malicious email will select malware FVs from Section~\ref{sec:apps}
as input to their corresponding LDs, while the remaining nodes select benign
FVs.  



To simulate the infection spreading over the email network, we need to (a)
model when a recipient `opens' the email:
we do so using a long tail distribution of reply times where the median open
time is 47 minutes, 90-percentile is one day, and the most likely open time is
2 minutes~\cite{email-resp-dist}; and (b) model the `click' rate (probability
that a recipient clicks on a URL): we vary it from 0\% up to 100\% to control
the rate of infection.
For example, within 1-, 2-, 3-hour long time interval only 55\%, 65\%, and 70\%
of recipients of a malicious email open it, which corresponds to 55, 65, and 75
infected machines respectively at 100\% click rate.

Overall, these two scenarios differ in their time-scales (seconds v. hours)
and in the relative rate at which benign and malicious neighborhoods grow.
As we will see, these parameters have a significant impact on the composition
of neighborhoods and the Shape GD's detection rate.



\begin{figure}[tbp]
   \vspace{-0.0in}
   \centering
   \includegraphics[width=0.45\textwidth]{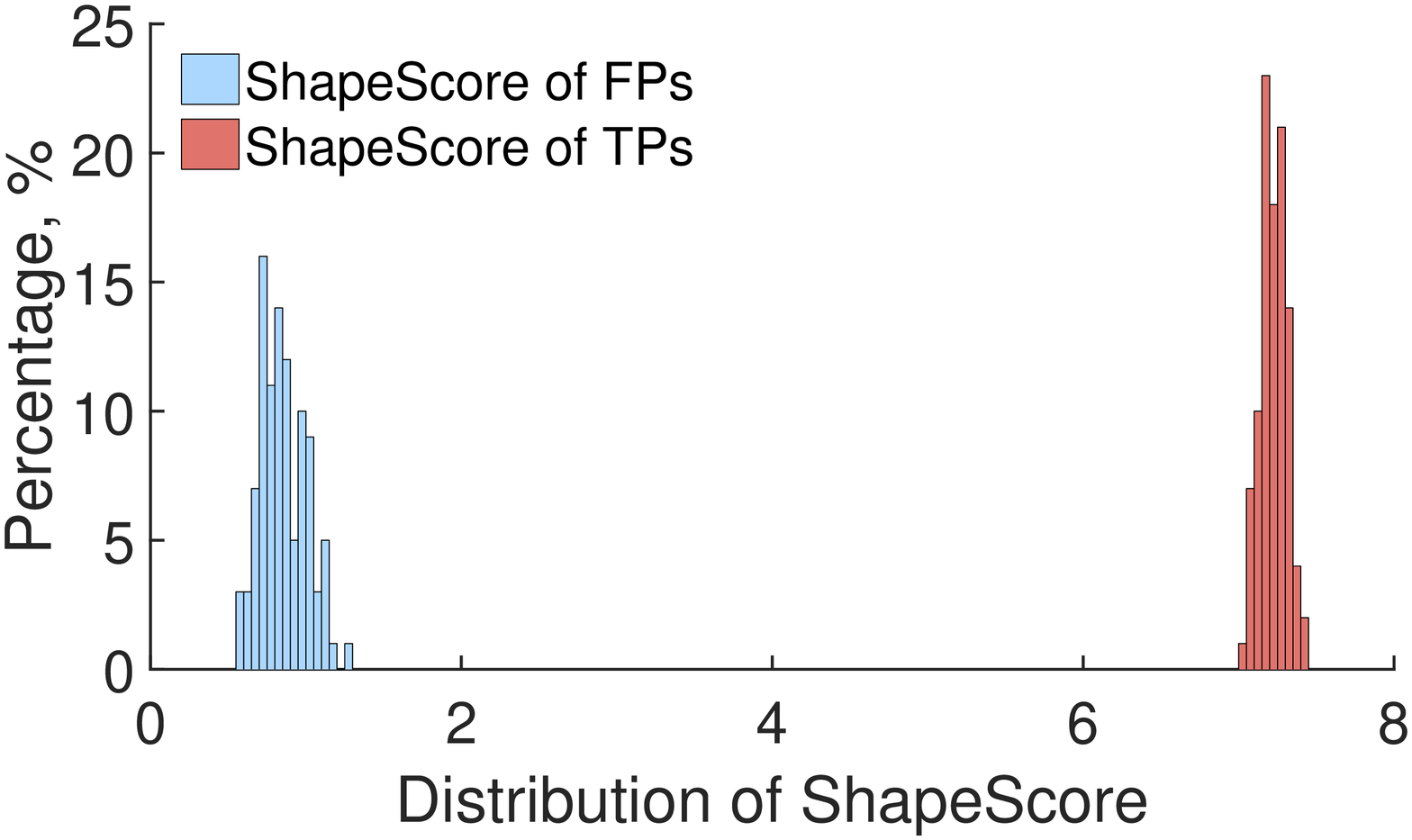}\par
   \caption{{\em Histogram of the ShapeScore: The ShapeScore is computed   	
     for neighborhoods with 15,000 FVs each (experiment repeated 500
     times to generate the histograms).  Shape-based GD can reliably
     separate FPs and TPs through extracting information from the data
     that has been unutilized by an LD.}}
   \vspace{-0.2in}
\label{fig:windows-hist}
\end{figure}

\noindent{\bf Methodology.} We report averaged results from repeating each
experiment multiple times with random initialization parameters.  In
particular, we use 10-fold cross validation for machine learning experiments
(Figure \ref{fig:windows-LD-roc}), 500 randomly sampled benign/malicious
neighborhoods with 10 repetitions to compute average (Figures
\ref{fig:windows-hist}, \ref{fig:windows-hist-stability}), 100 repetitions of
each malware infection experiment (Figures
\ref{fig:windows-shape_vs_count},\ref{fig:windows-shape_vs_count-waterhole},\ref{fig:windows-noisy_count},\ref{fig:windows-noisy_count-waterhole}),
and 100 repetitions of infection with 10 repetitions per data-point (Figures
\ref{fig:windows-time-NF-phishing},\ref{fig:waterhole_time_NF},\ref{fig:windows-detection_vs_censoring_threshold},\ref{fig:waterhole-structural-NF}).
To train the reference histogram, $H_{{\rm ref}}$, we select
15K FVs and 100K FVs from the training data set in the phishing and waterhole
experiments respectively.
All Shape GD's parameters are chosen based on a training data set (used for
Figures~\ref{fig:windows-hist} and~\ref{fig:windows-hist-stability}) -- we then
evaluate Shape GD (in the remaining figures) using a completely separate
testing data set.  

 



\ignore
{
\subsection{Virtual Machine Experimental Setup}
\label{sec:impl}
For our large-scale experiments, we deploy our data collection system on Amazon Web Services (AWS).
Experiment nodes run Windows Server 2008 R2 Base on default T2 micro instances with 1 GiB RAM, 1 vCPU, and 50 GB local storage. 
We populate the experiment environment with user data commonly found on a real host including: 
PDFs, word documents, photos, Firefox browser history, Thunderbird calendar entries, and Thunderbird contacts.
We then create a snapshot image of the virtual machine environment and clone this image for each experiment.
For simplicity, all malware experiments are conducted with a fresh clone of the base image.

Client code runs on-host and runs each experiment and communicates results with the central database.
Each experiment is run via a custom Intel Pin tool, which launches the target applications and intercepts system calls. 
This tool records timing information and system calls with arguments, and is configured to follow spawned child processes.
In comparison to full system emulation approach, Pin incurs moderate overhead while transparently instrumenting binaries.

The worker instances run in a virtual private cloud (VPC).
We simulate a local network via two subnets in the VPC. 
\textit{Neigboring machines} exist in a private subnet with no internet access, but are accesible by the experiment machines.
If malware happens to propogate to one of these vulnerable machines then it is, by default, quarantined.
Experiment machines run on the second subnet with limited public internet access. 
The second subnet allows malware to communicate over common C\&C ports and the client code to post results.
Specifically, we allow access for HTTP, HTTPS, SMTP, DNS, Telnet, and IRC.


\subsection{Benign Applications}

\begin{table*}[t]
\centering
\begin{tabular}{ p{1.5cm}  p{3.3cm} p{1.2cm} p{1.5cm} p{1.5cm} p{1.5cm} p{1.5cm} p{1.25cm}}
  \multicolumn{8}{c}{User Traces} \\
  \hline
  App & Description of User Activity & Exec time (min) & File Ops  & Network Ops & Process Ctrl & System Mgmt & Misc \\ \hline
  Wikipedia & Browse and search multiple articles & 480 & 1723054 &124003 &36649 &1508412 &5 \\            
  Allrecipes & Search for recipes, scroll through recipe photos & 510 & 24598084 & 11515472 &60076 &4757467 & 0 \\               
  Yelp & Browse local catering, read reviews & 500 &4270481 &949696 &178299 &4685311 & 28\\                       
  HTML5 Reference & Repeatedly open and read documentation & 545 &1745087 &264893 &110896 &4208092 &654 \\  
  \hline
  & \\                                          
  \multicolumn{8}{c}{Monkey} \\                \hline
App & Description of App & Exec time (min) & File Ops  & Network Ops & Process Ctrl & System Mgmt & Misc \\ \hline
  Walmart & Online shopping & 540 &4361902 &324263 &50866 &3937191 & 0\\                  
  AVG & Free antivirus & 800 & 16339990&826870 &245933 &8317314 &634 \\                       
  Buzzfeed & Social news & 520 &15733289 &1100059 &154094 &9380923 &0 \\                     
  eBay & Online Auction & 510 & 2178359&307148 &58990 &4716632 &0 \\                      
  Zedge & Wallpaper and Ringtones & 675 & 3779037&137739 &79588 &6052683 &3 \\            
  Holy Bible & Holy Text & 840 & 1229925&761161 &97201 &10703697 &10 \\                 
  \hline \\
\end{tabular}
\caption{Characterization of Applications. All applications chosen from Google Play Store top 100 free apps. Here, system calls are grouped into logical categories. For example, the \texttt{mount} system call is categorized as system-wide, and the \texttt{timer\_create} system call is classified under misc. }
\label{table:benign}
\end{table*}

Our platform currently includes 10 benign apps chosen from the top 100 Android
apps from the Google Play store.  
Table~\ref{table:benign} shows the apps and their aggregate system calls -- it
is clear that the chosen apps are popular as well as diverse in each system
call category.
In particular, we included apps such as online shopping clients (Amazon,
Walmart, eBay), anti-virus software (AVG) and a wallpaper app (Zedge) that are
popular carriers of malware, an entertainment news app (Buzzfeed), an
encyclopedia (Wikiepdia), online advertisement (Yelp), HTML5 reference app, and
a Bible. 


For each benign application, we supplied realistic and diverse user input to
observe variety of its dynamic behaviors. For this purpose we relied on two
different tools: TestDroid~\cite{testdroid} and Android Monkey
tool~\cite{adb-monkey}. The former can be used to record {\em real user input}
as a sequence of UI events and then to replay it multiple times preserving time
intervals between subsequent events. Monkey randomly generates user input
events such as touch, scroll, push events. We intentionally did not use
automatic input generation tools \cite{dynodroid,concolic-android} to explore
app's state space because they are still much worse than real user input and do
not work with custom app's layout.

We arranged 20 human participants to record dozens of 3-min long
interactions with four apps using TestDroid. 
Humans can produce a semantically meaningful sequence of inputs
(e.g. to log-in, satisfy ordering constraints when filling out a sequence
of fields in forms, etc).
Later, we replayed
these recordings multiple times to collect system call data. 

However, TestDroid has several limitations -- it often fails to replay UI sequences, especially when an app
extensively uses Android WebViews, 
or frequently causes an app under analysis to
crash -- and
we were unable to expand this approach to the other apps. 
Hence, we chose six other popular apps whose realistic usage can be closely
approximated by inputs generated by Monkey.

\noindent {\bf Realistic user data.} 
We populate our emulators with synthetic data that is statistically close to
the real user data, following data set sizes estimated by Kazdagli et
al~\cite{morpheus-hasp}. Without user data such as Contacts to steal, 
even a functional information stealer malware may not execute 
malicious code.

\label{sec:benign}

\subsection{Diverse Malicious Behaviors}


We evaluate our GD algorithms not simply against off-the-shelf malware samples,
but against malware samples that are guaranteed to execute correctly and have a
computationally diverse payload.
We now describe the range of malware behaviors and our process to construct a
parameterized, state-of-the-art malware sample.


\noindent {\bf Limitations of off-the-shelf malware.} Most malware samples
available
online~\cite{dissect_malware_2012,contagiodump,malware.lu,virusshare.com} do
not execute correctly. This is because malware may require older, vulnerable
versions of Android OS; they are designed to run only in a specific
geographical location; include anti-emulation
defenses~\cite{anti_emulation_android_vidas}; passively wait for commands from
command and control (C\&C) servers that are temporarily or permanently down;
react to specific user actions.  Another challenge of using off-the-shelf
malware is that
malware code-names assigned by anti-virus (AV) companies -- like
\textit{CruseWin} and \textit{AngryBirds-LeNa.C} -- do not inform an analyst
what payload the malware actually executes. Hence we evaluate our detector
against malware that we have reverse-engineered and ensured to work correctly
(on both the device and C\&C server sides); whose payloads we understand in
behavioral terms (such as stealing files or Contacts); and whose payloads we
actively diversify to stress test our detector.




\noindent {\bf Executing diverse malicious behaviors.} 
We analyze 229 malware samples from 126 families drawn from public
repositories~\cite{contagiodump,malware.lu,virusshare.com} (dating from
2012--2015).  We identify common malware payload behaviors and design patterns 
and extend a Geinimi.a sample~\cite{geinimi} to implement these payloads in a
parameterizable manner.

%
\begin{table*}[tb]
\centering

\begin{tabular}{p{3cm} >{\raggedright\arraybackslash}p{2.25cm} p{1.75cm} >{\raggedright\arraybackslash}p{1.5cm} >{\raggedright\arraybackslash}p{2cm}}
Syntetic Malware & Parameters (per payload action) & Malware Spec. Delay (ms) & Exec. Time (min) & Num. Syscalls (Million)\\ 
\hline

File Stealer (4.2MB each) &1, 10, 50 &0, 5k & 3215 & 493.8\\
Contact Stealer & 25, 250 &0, 50 & 1520 & 238.7\\
SMS Stealer & 50, 1000&0, 50 & 1525 & 236.4\\
ID, GPS Stealer &data size fixed &0, 200 & 540 & 89.5\\
Click Fraud (webpages) & 20, 300&0, 3k & 1490 & 223.7\\
DDos (slow loris) & 500 Connections & 1, 200 & 495 & 73.9\\
SHA1 passwd. Cracker & 10k, 1.5M&0, 50 & 1540 & 242.3\\
\hline \\
\end{tabular}
\caption{Summary of synthetic malware payload configurations and resulting system call traces. Note, the total execution times and system call counts were counted only for the one-second intervals where malware payload was active.}
\label{table:malware}
\end{table*}

We find that malware payloads fall into {\em three} orthogonal behavioral
categories:
{\em information stealers, networked nodes, and compute nodes}. C\&C server may
instruct client-side code to execute any combination of atomic tasks drawn from
these categories. 


\noindent {\bf 1. Information stealers.} Malware stealing personal information
usually focuses on accessing the following data: contacts for spamming
purposes, text messages for breaking two-factor authentication (e.g. bank
trojans), location and files on device for spying on users, phone IDs for
legitimizing stolen devices.  The Geinimi.a sample we start from already
implemented commands to execute these behaviors -- we add the tasks' intensity
and volume as parameters as shown in Table~\ref{table:malware}.

Contact stealing service requires two parameters: the number of
contacts being exfiltrated and delay between sending queries to the contact
provider. SMS stealers in real malware samples act either as a batch
stealer or as an intelligent stealer (e.g. bank trojans). 
Batch stealers transmit all the text
messages to their C\&C server, while the latter register an Android listener to
intercept incoming SMS messages and scan them for the presence of
authentication codes. Alternatively, they can perform intelligent text search
within the database of already received messages and upload 
only sensitive ones to the C\&C server.

To steal location information in Android, our malware can be configured to
register an Android listener that receives location updates or request location
directly from the Android middleware.  In the latter case, Android returns a
previously cached location that may have been determined using cellular network
(approximate mode) or GPS (precise mode) if it is available on a device.

Our extended Geinimi.a malware supports all file operations: it can steal
directory contents, upload a particular file or several files to the C\&C
server, or download a file received from a server to the device.  The
download-file functionality is widely used by malware developers to install new
(likely malicious) apps.

Finally, malware steals device IDs such as IMEI and IMSI codes, OS info
available via the static class android.os.Build, and other miscellaneous device
specific data such as browsing history and bookmarks, description of installed
apps, call logs. This information, especially IMEI and IMSI IDs, is used to
impersonate other devices --- specifically, to legitimize stolen devices before
selling them on a black market and for targeted advertising.

In summary, our Geinimi.a sample allows an analyst to specify the
following parameters: amount of data being read (e.g. 1 or 10
contacts/SMSs/files), intensity of data accessing operations (e.g. read a file
by 1K chunks), delay between successive reading operations (e.g. sleep for 100
ms between retrieving subsequent contacts/SMSs).

\noindent {\bf 2. Networked nodes.} 
This category exploits the phone as a device on a network
of nodes, e.g., to run click fraud or distributed denial-of-service (DDoS) 
attacks~\cite{malware_blackhat_15, prolexic_q4_2013,
android_ddos_2012}. 
Mobile devices are particularly attractice since their 
IP changes frequently, making it hard to blacklist one.

Click fraud is a popular, revenue-generating payload. The malicious service
receives a list of URLs from a C\&C, periodically fetches webpages specified in
the URL list and traverses their DOM structure. To speed up the process, our
Geinimi.a extension can be configured to launch several parallel threads.
A majority of DDoS attacks abuse the HTTP protocol (e.g. 80\%
according to \cite{ddos_threat_spectrum_2012}). Our malware is able to mount
two attacks: GET flood and SlowLoris~\cite{slowloris_ddos}. GET flood comprises
a series of GET requests sent by compromised network nodes. As opposed to GET
flood, SlowLoris attack is far less computationally expensive because it tries
to exhaust server pool of available connections by opening numerous connections
and sending data very slowly to keep connections alive over a long time period.
In our experiments, synthetic malware establishes 500 connections. If some
connections fail, they are automatically reopened to keep the total number of
active connections constant.

\noindent {\bf 3. Compute nodes.} The last category of malicious behaviors
includes computationally intensive malware. Unlike the previous two, it may or
may not leave distinguishable system-call fingerprint. A typical
example of such malware is a bitcoin miner \cite{mobile-bitcoin-miner}. We
approximate this category by the code cracking SHA1 passwords.

\ignore{
\noindent {\bf Realistic user data.} 
We populate our emulators with synthetic data that is statistically close to
the real user data, following data set sizes estimated by Kazdagli et
al~\cite{morpheus-hasp}. Without user data such as Contacts to steal, 
even a functional information stealer malware may not execute 
malicious code. 
}

\noindent {\bf Repackaging Android apps with malware payloads.} 
We embed malware payloads
into benign underlying apps using the same methods as malware developers:
disassembling an apk using \textit{apktool}, copying malicious code into the apk,
and modifying \textit{Manifest.xml} to extend the list of required permissions and
to statically register malicious components. Finally, we reassemble the
decompiled app with \textit{apktool} and sign it using \textit{jarsigner}.

\ignore{
\noindent {\bf Ensuring correct execution.} One of our key design decisions 
is to augment malware to notify the platform 
when events happen (e.g. C\&C request has been successfully
parsed, malware starts/finishes execution). This is very important because
android emulators often crash, thus we can identify and rerun failed
experiments. We also use malware start and stop time stamps to start and stop
collecting malicious feature vectors.  We also make sure that the notification
mechanism does not pollute the collected data.
}

Malware payload on a device perform malicious actions in response to the
commands that it receives from its C\&C server and communicates back to the
server to confirm successful completion.  For example, the service performing
click fraud activity creates a thread pool and supplies a list of URL links to
the workers in the pool that they must access.  Most Android apps (both benign
and malicious) are obfuscated using standard Android tool --
Proguard~\cite{proguard} -- which renames classes, fields, and methods with obscure
names. We have applied Progaurd to our modified Geinimi.a malware as well.

\noindent{\bf Diversity of malware payloads.} Table~\ref{table:malware}
shows the parameters used for each payload category 
and the effect of payloads on system call traces (in aggregate). It is clear
that the payloads differ in system call intensities and (as we find in practice)
yield a diverse set of machine learning features.


\ignore{When a client receives a
notification from C\&C in the form of an xml file, it parses server's request,
configures itself and launches one or more Android services responsible for
carrying out a malicious task. Depending on C\&C's request, malware may run
those services sequentially or concurrently. Some services act as an
intermediate dispatcher -- they perform initialization and then spawn parallel
threads to conduct actual malicious activity asynchronously. For example, the
service performing click fraud activity creates a thread pool and supplies a
list of URL links to the workers in the pool that they must access.
}



\label{sec:synthetic_malware}



\subsection{Local Detectors and Feature Extraction algorithms}
\label{sec:LDs}
\begin{figure*}[t]
  \begin{minipage}[tbp]{0.33\linewidth}
    \centering
    \includegraphics[width=\textwidth]{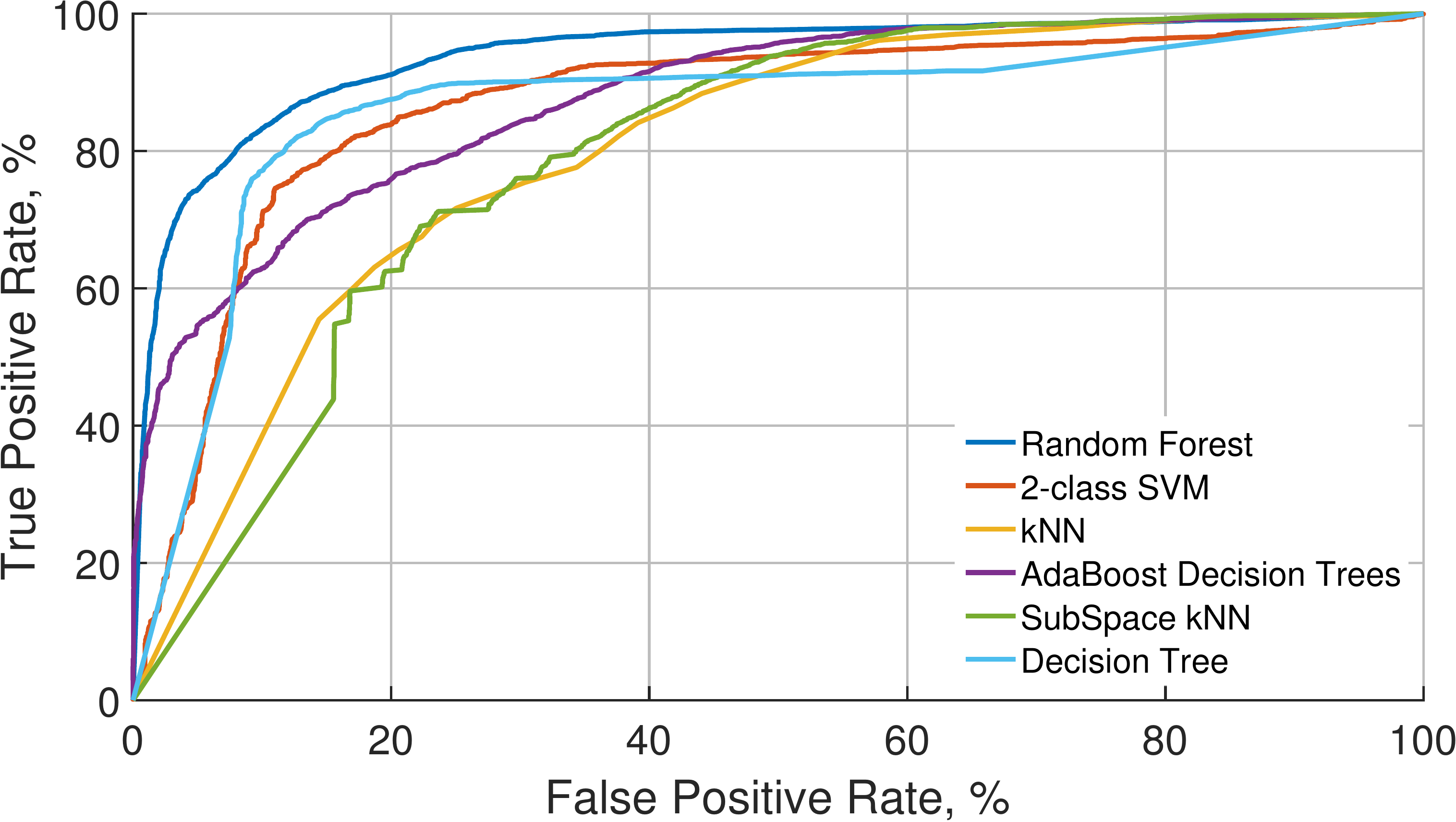}\par
    \subcaption{Histogram FE algorithm.}
    \label{fig:roc_pca_wiki}
  \end{minipage}
  \begin{minipage}[tbp]{0.33\linewidth}
    \centering
    \includegraphics[width=\textwidth]{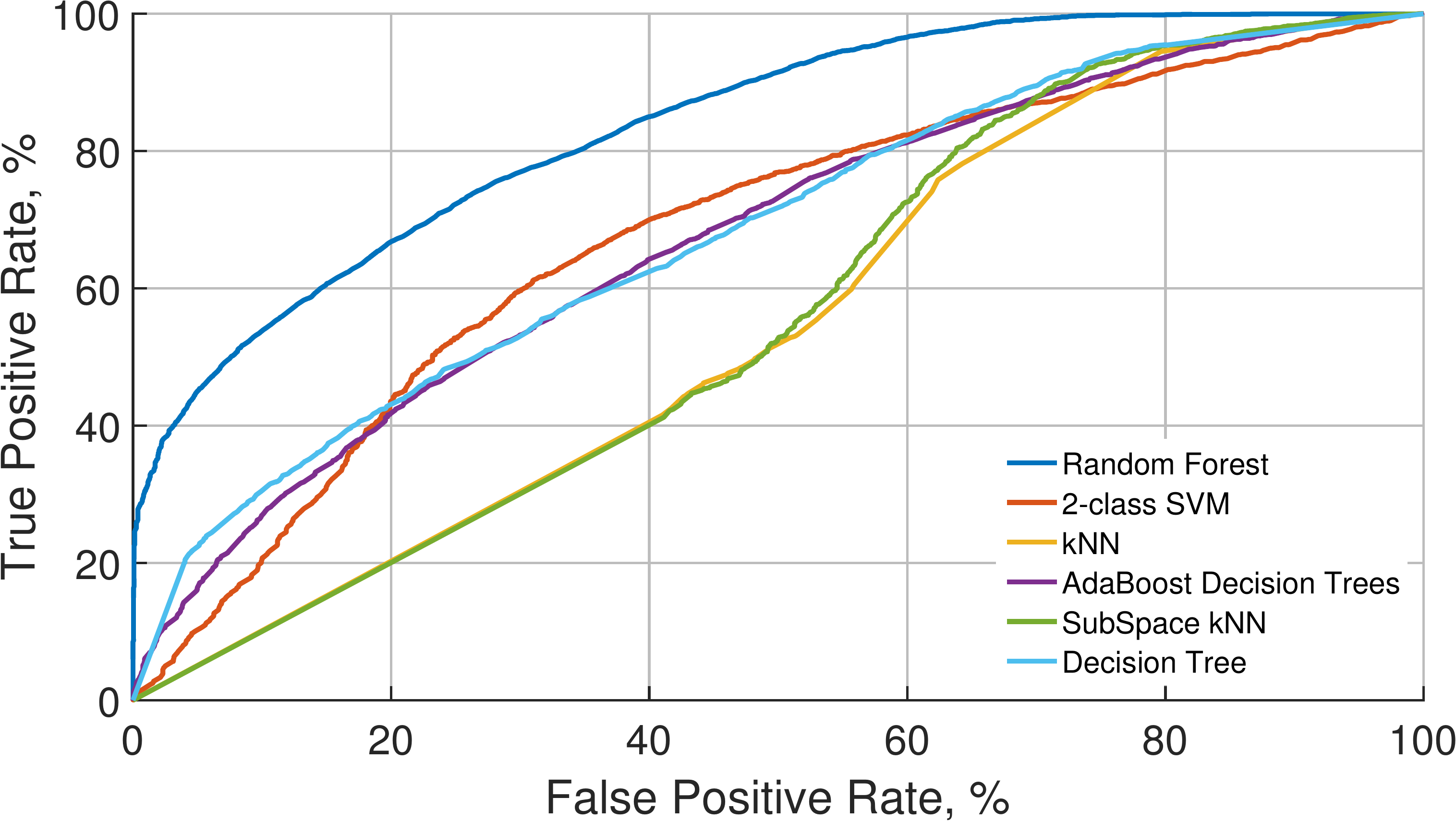}\par
    \subcaption{N-gram FE algorithm.}
    \label{fig:ROC_ngram_wiki}
  \end{minipage}
  \begin{minipage}[tbp]{0.33\linewidth}
    \centering
    \includegraphics[width=\textwidth]{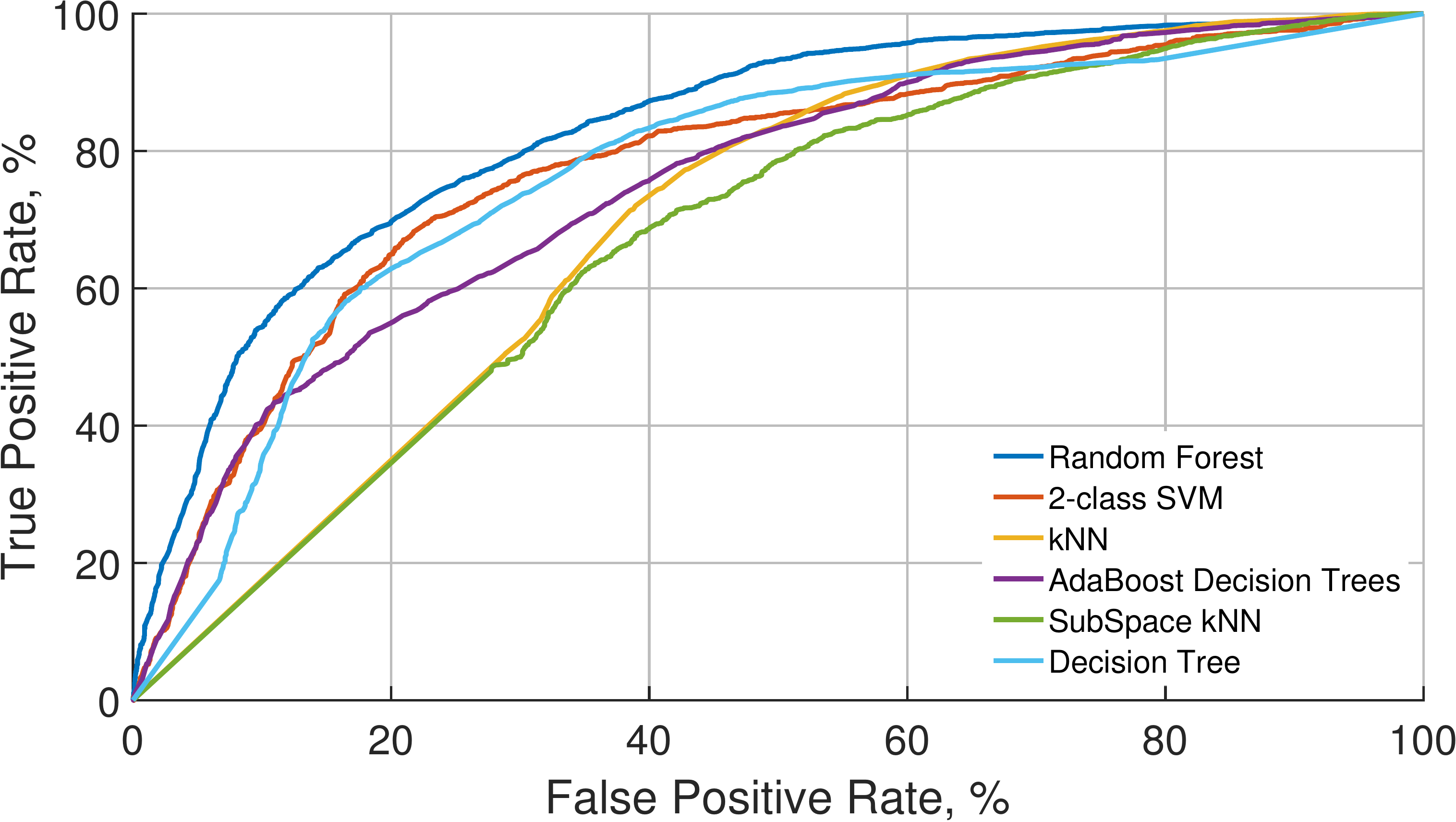}\par
    \subcaption{Histogram of manually classified system calls.}
    \label{fig:ROC_intuitive_basis_wiki}
  \end{minipage}
  \caption{
(Wikipedia app) True positive v. False positive (ROC) curves shows
    detection accuracy of six  local detectors. In terms of detection accuracy
    and computational efficiency, Random Forest together with histogram feature
    extraction (FE) algorithm outperforms the rest classifiers and FE
    algorithms.
    However, any of these local malware detectors has unacceptably high false
    positive rate if one wants to achieve at least 80\% TP rate.}
  \label{fig:ROC}
\end{figure*}


Our first step is to establish a good local detector (LD) for Android devices.
In particular, we choose system call based LDs since the system call interface
has visibility into an app's file, device, and network activity and can thus
capture signals relevant to malware executions.  Interestingly, 
unlike system call detectors for Linux and Windows, detectors on mobile
platforms like Android have not been evaluated by prior work.  
Doing so is important since
Android adds several middleware services that consolidate 
system call traces traces from third-party applications -- for example, services that 
manage
Contacts, SMSs, sensors, etc. Further, Android also 
provides a different event-driven
programming model and software stack compared to Windows and Linux. 

We experiment with an extensive set of system-call LDs -- our takeaway is that
even the best LD we could construct operates at a true- and false-positive
ratio of 80\%:15\% and is not deployable by itself.

\ignore{
The high false positive rate is likely due to the nature of mobile malware
as well as our evaluation method that emphasizes realistic usage of benign 
applications.\mohit{point to experimental setup here}
Note that alternative LDs based on
hardware~\cite{Demme2013,ponomarev-hpca-15} or Android middleware~\cite{Enck2009,Peng2012} also report
similarly high false positives -- all such LDs' results can be improved using
our proposed Global Detector (GD).
Further, improvements to LDs are orthogonal to \sysname and only serve to
further improve the time to detect a malware outbreak.
}

Each LD 
comprises of a machine learning (ML) classifier and a feature extraction (FE)
algorithm. 
We choose LD candidates from general machine learning (ML) models that are
computationally efficient to train -- such as SVMs, random forest, k-Nearest
Neighbors, etc and not including the more complex artificial neural networks or
deep learning algorithms.  We deliberately avoid handcrafted ML algorithms and
hardcoded detection rules because such custom models usually are overfitted to
a specific dataset, and because they require humans to continuously adjust the
detector to new types of malware.

To extract features, we experiment with
histograms, n-grams, and also using a human-understandable system call basis.
When using histograms and n-grams, we compute the (top ten) principal
components using PCA analysis to reduce dimensionality of the data. 
For the human-understandable \intbasis, we group system calls into
33 dimensions each related to file access, networking, process control, etc.
We use \intbasis as a candidate because it enables malware analysts to 
map malware payloads back to intuitive tasks such as stealing SMSs or snooping
on location data.

Using these features, we experiment with six state-of-the-art ML algorithms: random
forest, 2-class SVM, kNN, decision trees, and their ensemble versions --
boosted decision trees with AdaBoost algorithm and Random SubSpace ensemble of
kNN classifiers~\ref{fig:ROC}. We also evaluated 1-class SVM as an anomaly
detector and Naive Bayes classifier -- however, both yielded an extremely high
FP rate and we exclude them from further discussion.

In total, we consider these 18 potential LDs (three FE algorithms x six ML
classifiers) to choose the best one.  We compare all of them using ROC
curves~\ref{fig:ROC} that plot true positive v. false positive rates, so that
the area under the ROC curve (AUC) is a quantitative measure of LD's
performance: the larger the AUC, the more accurate the detector.
To construct ROC curves we apply 10-fold cross validation to avoid any bias
from the dataset, i.e. our classifier predicts labels of in-fold observations
using a model trained on out-of-fold observations. 

\subsection{Synthetic and Real Networks}

\ignore{
We now present our choice in representative graphs for describing malicious
activity on networks. We propose two naive synthetic networks,
Erd\H{o}s-R\'{e}nyi and grid graphs, and generate independent communities
connected with a known probability. For simplicity, each community contains
approximately an identical number of nodes.
This conservatively models the situation where we operate on a sub-graph of a
larger network with coarse grained community detection, which is the worst case
scenario for our detection system. In addition, we test on several real world
graphs taken from the SNAP database \cite{snapnets}.      
}

The first type of community graph is an Erd\H{o}s-R\'{e}nyi model with $|V|$
nodes and $|E|$ edges.  Given the size of the community as $A_c = \{ v \in V :
v \in c\}$ then we estimate the probability of intra-community connection
probability to be $p = 2*|E|/|A_c|^2$.  We choose $p$ such that the degree of
the network, $M$, is between three and four\footnote{Nodes with zero degree are
removed.} so that the network is not too dense, mimicking many realistic
graphs. We generate a random graph with 2 communities of (approximately) 1k
nodes such that the communities are equal in size. The resulting graph contains
958 nodes, 3108 edges, and approximately 600 nodes per community.

The second community graph we construct is a grid graph. For each community, we
generate a lattice-mesh with a known number of vertices and edges. 
We generate a grid with 2 communities,
1250 nodes, 4950 edges, and 625 nodes per community.  
  
Lastly, we use a subset of the Youtube graph from the Stanford SNAP database
\cite{snapnets}.  
We choose a sub-graph as follows: First, we pick a random node and find its
k-hop neighbors, building a sub graph $G'$ from these nodes to preserve
community structure; next we choose a random sub-graph from $G'$, removing all
the zero degree nodes until the desired number of nodes is reached.  Sub-graphs
picked in this way preserve the community structure of the network
Using this algorithm on the YouTube social network, we extract a subgraph of
1005 nodes, 2906 edges, and 19 communities.

\noindent{\bf Malware Spread Dynamics.}  The malware spreads according
to a standard SI model \cite{Durrett07}. In brief, the SI process is a
random infection process spreading on the graph. Each edge in the
graph has a time-interval associated with it (exponential, independent
across edges). Initially, all nodes are tagged `0' (uninfected),
except a source node which is tagged `1' (infected). The edge weights
(time intervals) now propagate this `1' throughout the network. The
time at which a specific node becomes `1' is simply the sum of the
weights along the shortest weighted path on the graph connecting the
source node to this specific node.



\subsection{Implementation Details}
For our large-scale experiments, we deploy \sysname on Amazon Web Services (AWS).
Experiment nodes run Windows Server 2008 R2 Base on default T2 micro instances with 1 GiB RAM, 1 vCPU, and 50 GB local storage. 
We Populate the experiment environment with user data commonly found on a real host including: 
PDFs, word documents, photos, Firefox browser history, Thunderbird calendar entries, and Thunderbird contacts.
We then create a snapshot image of the virtual machine environment and clone this image for each experiment.
For simplicity, all malware experiments are conducted with a fresh clone of the base image.

Client code runs on-host and runs each experiment and communicates results with the central database.
Each experiment is run via a custom Intel Pin tool, which launches the target applications and intercepts system calls. 
This tool registers timing information and system calls with arguments, and is configured to follow spawned child processes.
We chose Pin because it instruments binaries without modification in a format comparable to \tttext{strace} in Linux. 

Simulated local network
The worker instances run in a virtual private cloud (VPC).
We simulate a local network via two subnets in the VPC. 
\textit{Neigboring machines} exist in a private subnet with no internet access, but are accesible by the experiment machines.
If malware happens to propogate to one of these vulnerable machines then it is, by default, quarantined.
Experiment machines run on the second subnet with limited public internet access. 
The second subnet allows malware to communicate over common C&C ports and the client code to post results.
Specifically, we allow access for HTTP, HTTPS, SMTP, DNS, Telnet, and IRC.   

}

\vspace{-0.1in}
\section{Results}
\label{sec:results}
\ignore{use question and key findings in bold format}


%
We show that \sysname can identify malicious neighborhoods with {\em less than
1\% false positive and 100\% true positive rate when the neighborhoods produce
more than 15,000 FVs within a neighborhood time window (i.e., $|B| > 15,000$ in
Algorithm~\ref{ShapeGD_algorithm}}).  Recall that at 60 FVs/node/minute, it
takes 1000 nodes only 15 seconds to create 15,000 FVs. For LDs like ours with
$\sim$6\% false positive rate, this corresponds to 900 alert-FVs.  We then
simulate realistic attack scenarios and find that \sysname can detect malware
when only 5 of 1086 nodes are infected through phishing in an enterprise email
network, and when only 108 of 550K possible nodes are infected through a
waterhole attack using a popular web-service. Finally, \sysname is
computationally efficient -- we relegate this discussion to
Appendix~\ref{sec:overhead-appendix}.


\subsection{Can shape of alert-FVs identify malicious neighborhoods?}
\label{sec:power-shape}


We first show that the shape of a neighborhood can easily distinguish between
neighborhoods that are either 100\% benign or 100\% malicious.  
We quantify Shape-GD's time to detection under real settings with
a mix of both in subsequent sections.

Figure~\ref{fig:windows-hist} shows that \sysname can indeed
separate purely benign neighborhoods from
purely malicious ones. To conduct this experiment, we construct 
purely benign and malicious neighborhoods 
with $\sim$15,000 benign or malicious FVs respectively (i.e., $|B|$ is 15,000). 
In Appendix~\ref{sec:nbd-size}, we experimentally quantify the sensitivity
of \sysname to the number of FVs in a neighborhood ($|B|$) and find that
neighborhoods with more than 15,000 FVs lead to robust global classification.

For each neighborhood, we use the Random Forest LD to generate {\em alert-FVs}
and use \sysname to compute the neighborhood's ShapeScore using the alert-FVs
from the neighborhood.  In Figure~\ref{fig:windows-hist}, we plot histogram of
ShapeScores  
for 500 benign and malicious FVs each -- each point in the blue (or red)
histogram represents the ShapeScore of a completely benign (or malicious)
neighborhood.  Recall that a small ShapeScore indicates the neighborhood's
statistical shape is similar to that of a benign one.  {\em The non-overlapping
distributions separated by a large gap indicate that the shape of purely benign
neighborhoods is very different from the shape of purely malicious
neighborhoods.}


\sysname detects anomalous neighborhoods by setting a threshold score based on
the distribution of benign neighborhoods' scores (Figure
\ref{fig:windows-hist}) -- if an incoming neighborhood has a score above the
threshold, Shape GD labels it as `malicious', otherwise `benign'.  We set the
threshold score at 99-percentile (i.e. our expected {\em global false positive
rate} is 1\%) and the true positive rate is effectively 100\% for this
experiment.  This shows that for homogeneous neighborhoods producing over 15K
FVs within a neighborhood time window, \sysname can make robust predictions.
The next question then is how well \sysname can do so when neighborhoods are
partially infected -- we evaluate this in the next section.







\subsection{Time to detection using temporal neighborhoods}
\label{sec:time-nf}

Temporal filtering creates a neighborhood using only the nodes that are {\em
active} within a neighborhood time window (NTW). For example, a temporal
neighborhood for the phishing scenario would include every email address that
received an email within the last hour (1086 nodes in our experiments). 
Similarly, a waterhole attack scenario
would include all client devices that accessed {\em any} server within the last 
NTW into one neighborhood ($\sim17,000$ nodes on average in 30 seconds). 
This neighborhood filtering models a CIDS
designed to detect malware 
whose infection exhibits temporal locality (and obviously does not detect
attacks that target a few high-value nodes through temporally uncorrelated
vectors). 



Phishing and waterhole attacks operate at different time scales (and hence 
NTWs).  Due to the long tail distribution of email `open'
times, the phishing NTW varies from 1--3 hours in our experiments. On the other hand, a
popular waterhole server quickly infects a large number of clients within a
short period of time -- thus, we vary the waterhole NTW from 4 seconds up to 100 seconds.



\begin{figure}[tbp]
   \vspace{-0.0in}
   \centering
   \includegraphics[width=0.45\textwidth]{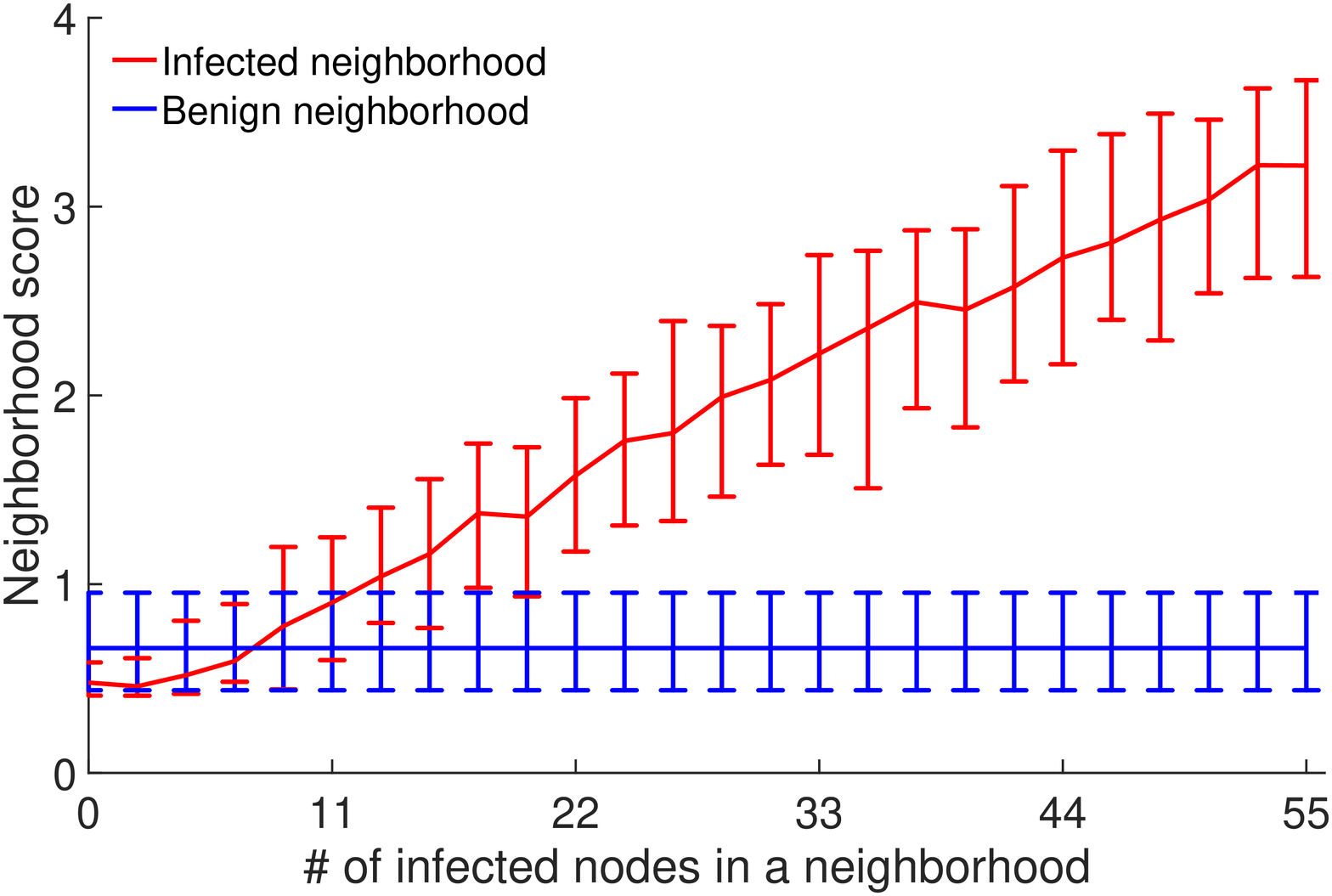}\par
   \caption{{\em (Phishing attack: Time-based NF) Dynamics of an attack:
     While the portion of infected nodes in a neighborhood increases
     over time reaching 55 nodes out of 1086 on average, ShapeScore
     goes up showing that Shape GD becomes more confident in labeling
     neighborhoods as `malicious'.  It starts detecting malware with
     at most 1\% false positive rate when it compromises roughly 22
     nodes.  The neighborhood includes all 1086 nodes in a network and
     spans over 1 hour time interval.}}
   \vspace{-0.2in}
\label{fig:windows-shape_vs_count}
\end{figure}

\noindent{\bf Shape GD's time to detection for one NTW.} 
We fix NTWs (1
hour for phishing and 30 seconds for waterhole) and vary
a parameter that represents a node's likelihood of infection from 0\% up to 100\% -- 
modeling whether a phished user clicks the malicious URL (phishing) or a 
drive-by exploit succeeeds in a waterhole attacks.   

\begin{figure}[tbp]
   \vspace{-0.0in}
   \centering
   \includegraphics[width=0.45\textwidth]{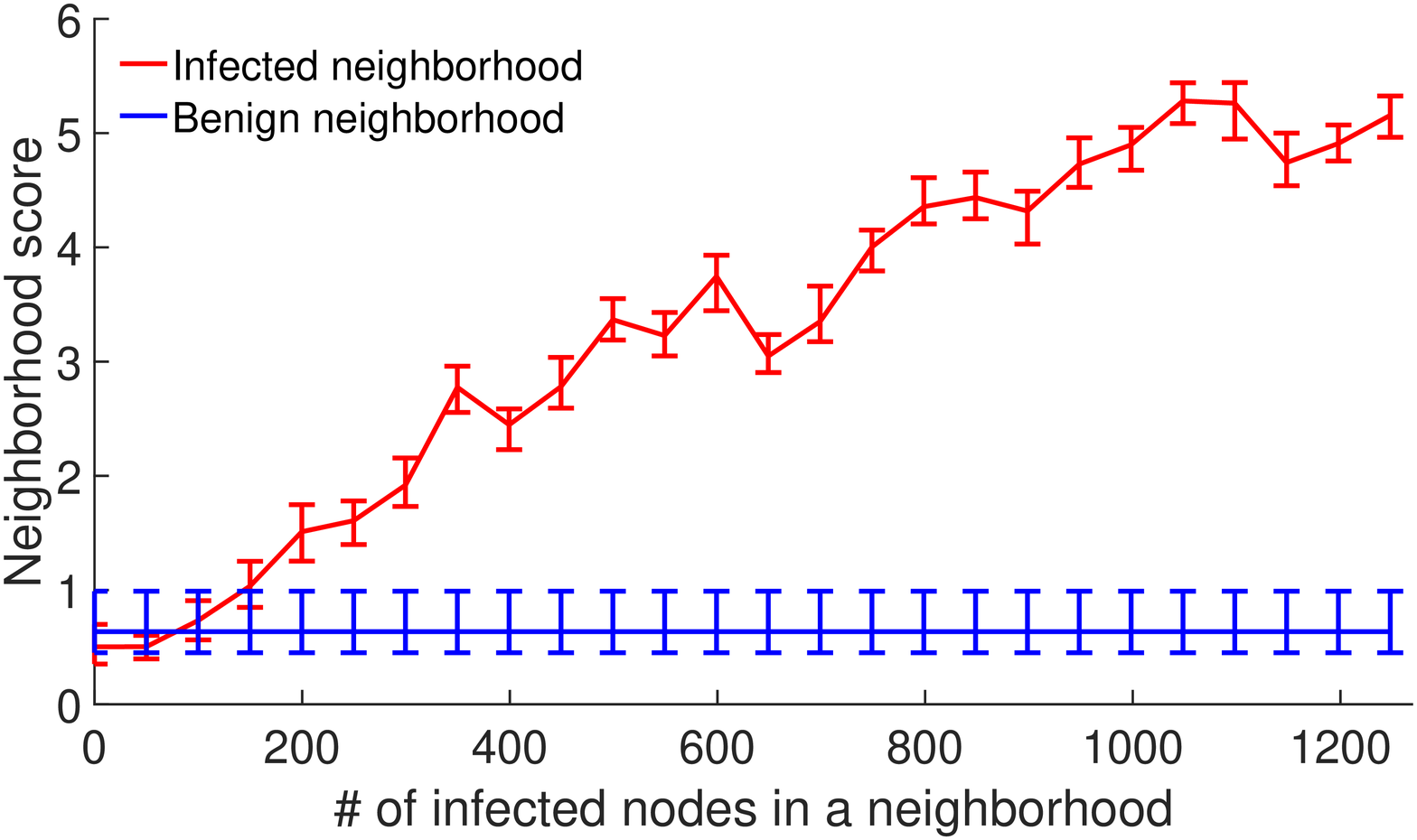}\par
   \caption{{\em (Waterhole attack: Time-based NF) Dynamics of an attack:
     While the portion of infected nodes in a neighborhood increases
     over time reaching 1248 nodes on average, ShapeScore goes up
     showing that Shape GD becomes more confident in labeling
     neighborhoods as `malicious'.  It starts detecting malware with
     at most 1\% false positive rate when roughly 200 nodes get compromised. 
     The neighborhood includes 17,178 nodes on average and
     spans over 30 sec time interval.}}
   \vspace{-0.2in}
\label{fig:windows-shape_vs_count-waterhole}
\end{figure}

Figures~\ref{fig:windows-shape_vs_count} and~\ref{fig:windows-shape_vs_count-waterhole}  
plot the neighborhood score v. the average number of infected
nodes within benign (blue curve) and malicious (red curve) neighborhoods 
-- the two extreme points on the X-axis corresponds
to either none of the machines being infected (the left side of a figure) or
the maximum possible number of machines being infected (the right side of the figure).  
In this experiment, phishing can infect up to 55 machines in the 1 hour
NTW, while the waterhole server can infect almost 1250 nodes 
in the 30 seconds NTW.
Every point on a line is the median neighborhood score from 10 experiments with
whiskers set at 1\%- and 99\%- percentile scores.


When increasing the number of infected nodes in a neighborhood, as expected,
the red curve larger deviates from the blue one.  Therefore, Shape GD becomes
more confident with labeling incoming partially infected neighborhoods as
malicious.  Shape GD starts reliably detecting malware very quickly -- 
when only 22 nodes (phishing) and 200 nodes (waterhole) have been infected.  
We also experimented with
other sizes of neighborhood window -- the plots we obtained showed similar
trends.
\ignore{min-max values for reliable detectio?}

\begin{figure}[tbp]
   \vspace{-0.0in}
   \centering
   \includegraphics[width=0.45\textwidth]{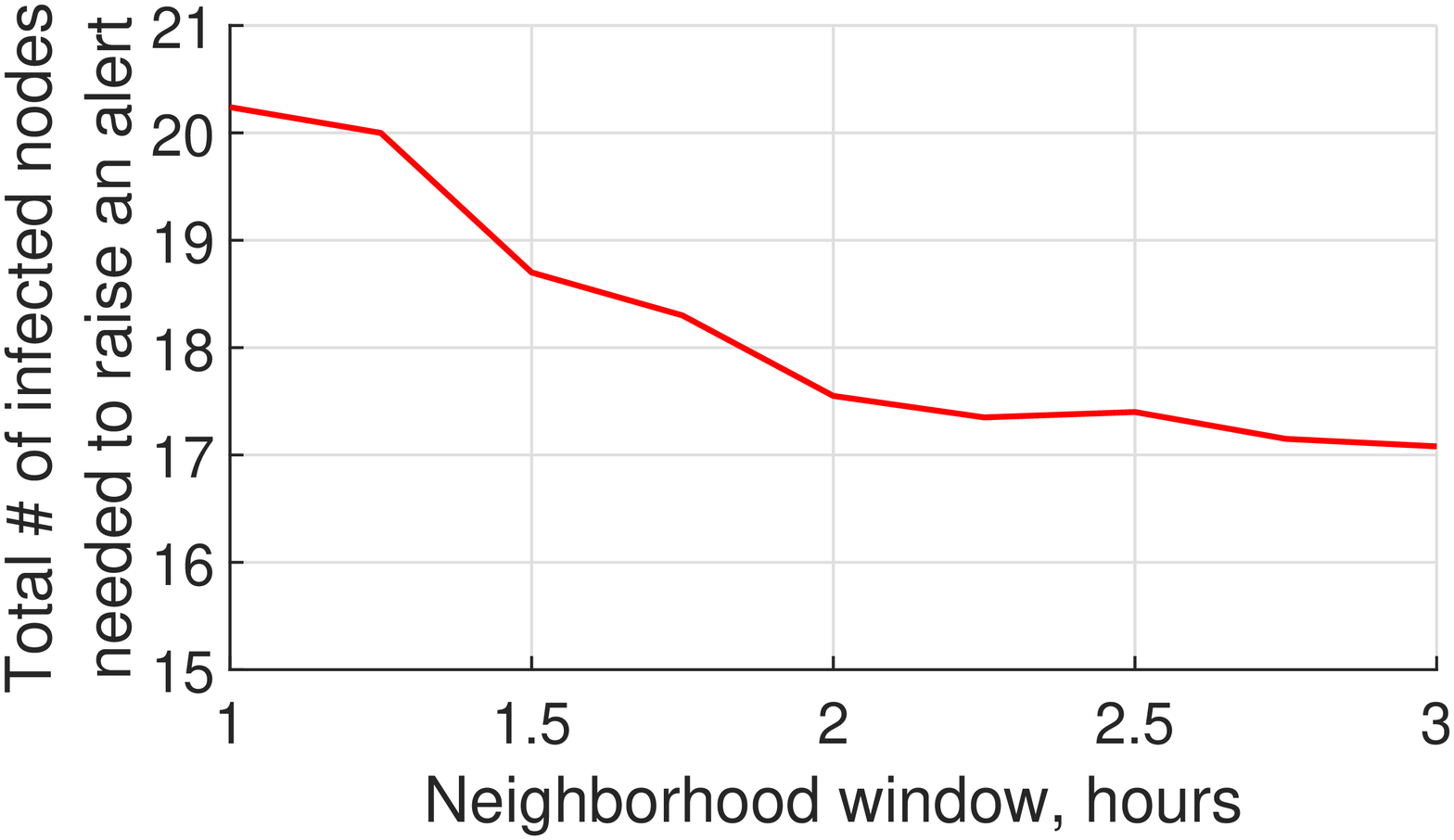}\par
\caption{{\em(Phishing attack: Time-based NF algorithm)  Shape GD's performance improves by 18.5\% (20.24 and 17.08 infected nodes) when increasing the size of a neighborhood window from 1 hour to 3 hours.
}}
   \vspace{-0.2in}
\label{fig:windows-time-NF-phishing}
\end{figure}

\noindent{\bf Shape GD's sensitivity to NTW.} We show that the size of a
neighborhood is important for early detection -- the minimum number of
nodes that are infected before Shape GD raises an alert -- in
Figures~\ref{fig:windows-time-NF-phishing} and~\ref{fig:waterhole_time_NF}.
Varying the NTW essentially competes the rates at which both malicious and
benign FVs accumulate -- interestingly, we find that \textit{these relative
rates are different for phishing and waterhole attacks and lead to different
trends for detection performance v. NTW}.

We vary the NTW from 1 hour to 3 hours for phishing and from 4 sec to
100 sec for waterhole and record the number of infected nodes when Shape GD can
make robust predictions (i.e. less than 1\% FP for almost 100\% TP).

Increasing the NTW in the phishing experiment from 1 to 3 hours
{\em improves} the Shape GD's detection performance -- at 17.08 infected nodes
for a 3 hour NTW compared to 
20.24 nodes for a 1 hour NTW.
Detection improves slowly because while the 
infection rate slows down over time as fewer emails remain to be
opened, the long tail distribution of email `open' times causes most of the 17
victims to fall early in the NTW and accumulate sufficient malicious FVs to tip
the overall neighborhood's shape into malicious category.  



In a waterhole scenario, the number of client devices active within a time
window (and hence the false positive alert-FVs from the
neighborhood) grows much faster than the malware can spread 
(even if we assume 
that {\em every} client that visits the waterhole server gets infected.
Here, a large NTW aggregates many more
benign (false positive) FVs from clients accessing non-compromised servers.
Hence, in contrast to the phishing attack,
increasing the NTW degrades time to detection. 
Shape GD works best with an
NTW of 6 seconds -- only 107.5 nodes on average become infected out of a
possible $\sim$550,000 nodes.
Note that a very small NTW (below 6 seconds) 
either does not accumulate enough FVs for analysis --
if so, Shape GD outputs no results --  or 
creates large variance in the shape of benign
neighborhoods and abruptly degrades detection performance.

Note that a Shape GD requires a minimum number of FVs per neighborhood to make
robust predictions -- at least 15,000 FVs based on
Appendix~\ref{sec:nbd-size} -- hence, the Shape GD has to set 
NTWs based on the rate of incoming requests and access
frequency of a particular server. For example, if a server is not very popular
and is likely to be compromised, the Shape GD could increase this server's NTW 
to collect more FVs for its neighborhood.

\subsection{Time to detection using structural information}
\label{sec:time-struct}



Both phishing and waterhole attacks impose a logical structure on nodes (beyond
their time of infection): phishing spreads malware through malicious email
attachments or links while waterhole attacks infect only the clients that
access a compromised server.  
This structure suggests that temporal neighborhoods can be further refined
based on the sender/recipient-list of an email (e.g., grouping members of a
mailing list into a neighborhood in the phishing scenario) or based on 
the specific server accessed by a client (i.e., grouping clients that visit 
a server into one neighborhood).

\begin{figure}[tbp]
   \vspace{-0.0in}
   \centering
   \includegraphics[width=0.45\textwidth]{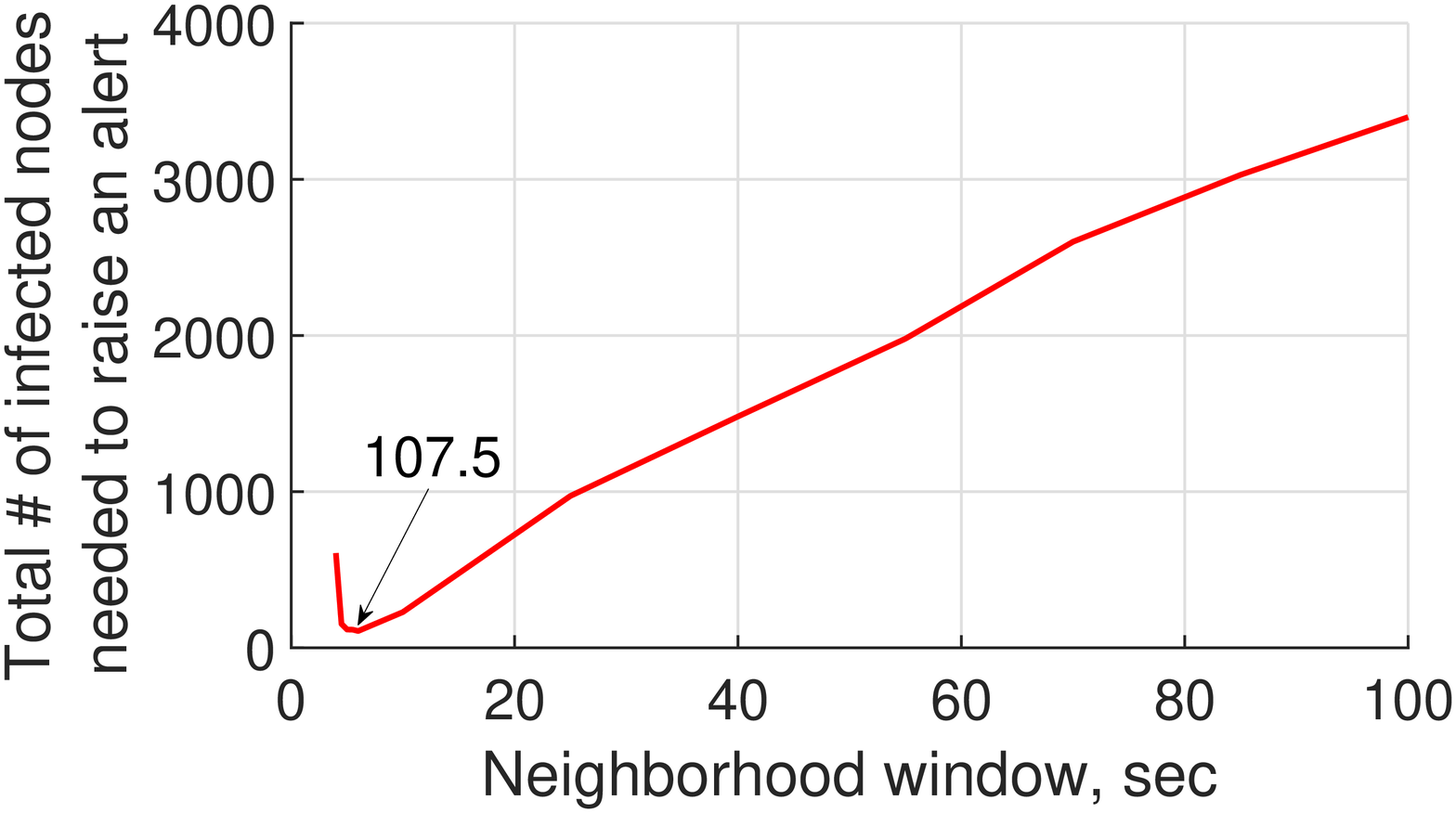}\par
\caption{{\em (Waterhole attack: Time-based NF) Shape GD's performance deteriorates linearly when increasing the size of a neighborhood window from 6 sec to 100 sec.
}}
   \vspace{-0.2in}
\label{fig:waterhole_time_NF}
\end{figure}

\begin{figure}[tbp]
   \vspace{-0.0in}
   \centering
   \includegraphics[width=0.45\textwidth]{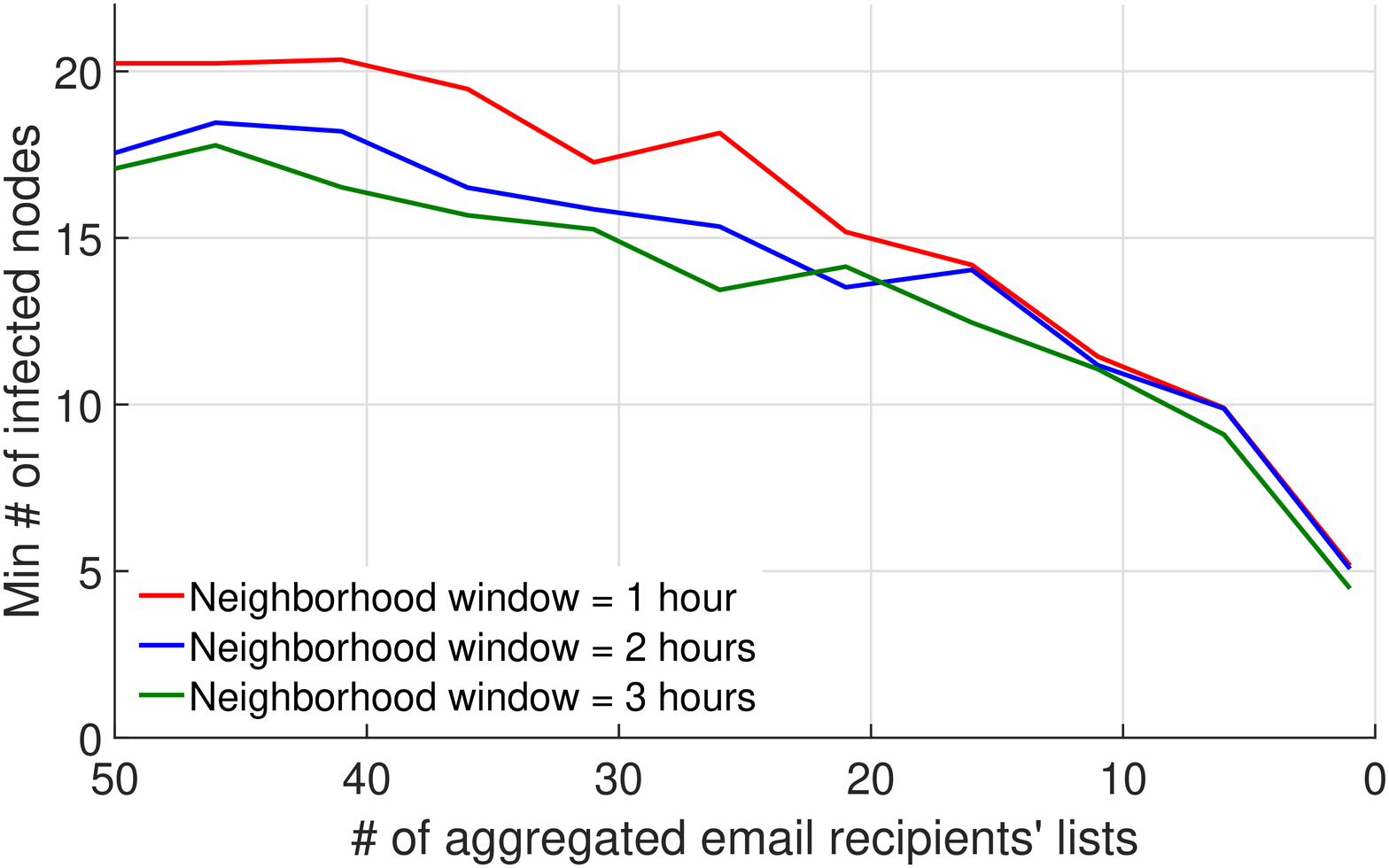}\par
   \caption{{\em(Phishing attack) Comparing to pure time-based NF,
     structural filtering algorithm improves Shape GD's performance by
     $\sim 4\times$ by taking into consideration logical structure of
     electronic communication (sender -- receiver relation).  }}
   \vspace{-0.2in}     
\label{fig:windows-detection_vs_censoring_threshold}
\end{figure}

\begin{figure}[tbp]
   \vspace{-0.0in}
   \centering
   \includegraphics[width=0.45\textwidth]{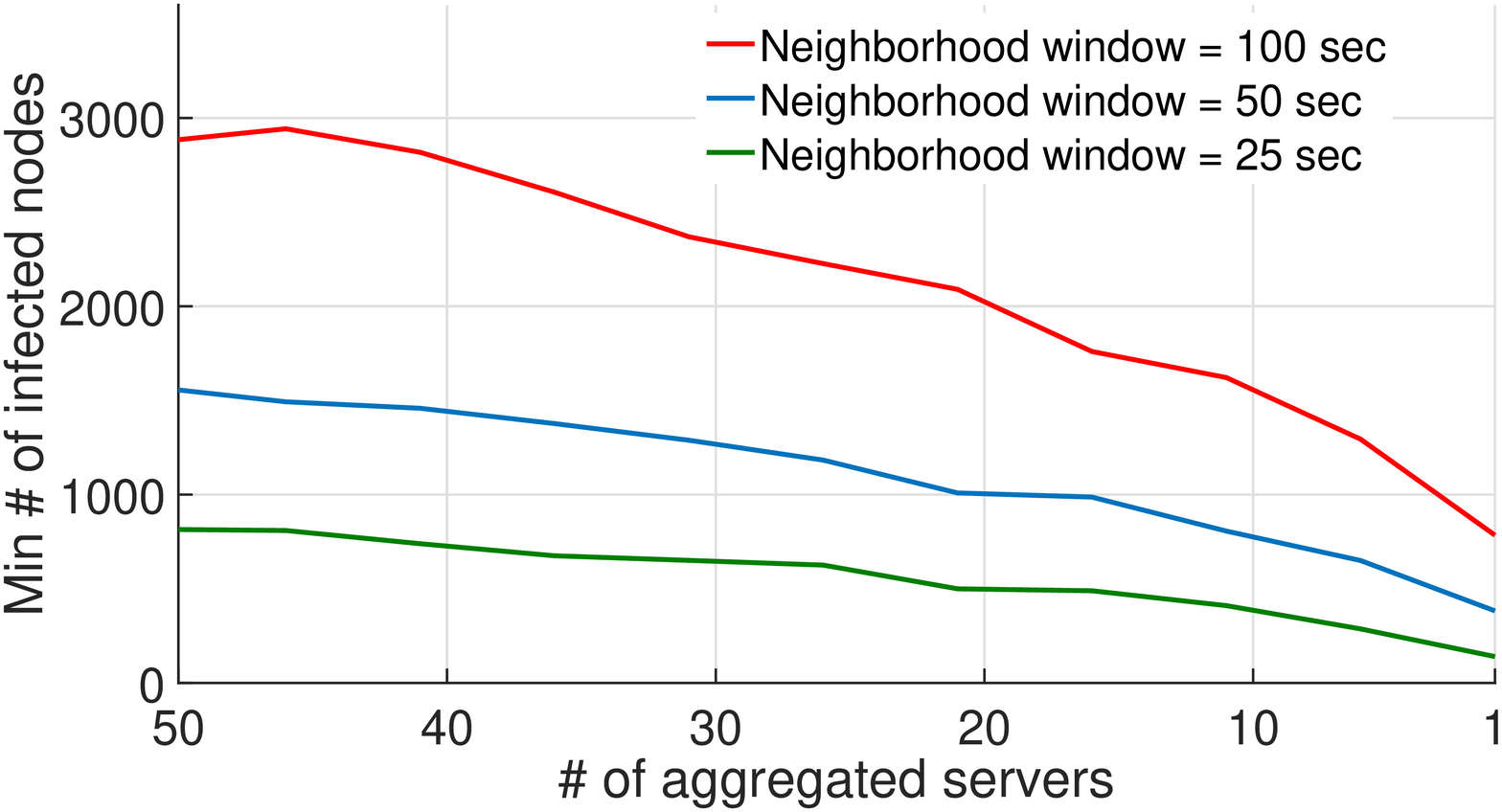}\par
   \caption{{\em(Waterhole attack) Comparing to pure time-based NF,
     structural filtering algorithm improves Shape GD's performance by
     $3.75\times$ -- $5.8\times$ by aggregating alerts on a server basis.  }}
   \vspace{-0.2in}     
\label{fig:waterhole-structural-NF}
\end{figure}

To analyze the effect of such structural filtering on GD's performance, we vary
filtering from coarse- (no structural filtering, only
time-based filtering) to fine-grained (aggregating alerts across each
recipients' list separately or across clients accessing each server separately)
(Figures~\ref{fig:windows-detection_vs_censoring_threshold},
\ref{fig:waterhole-structural-NF}).  Specifically, the aggregation parameter
changes from 50 recipients' lists or servers down to 1. \ignore{why 50? link back
to experimental setup.}
As before, we measure detection in terms of the minimum number of infected
nodes that lead to raising a global alert.  Also we consider three NTW values
-- 1-, 2-, and 3- hours long for phishing and 25-, 50-, and 100-sec long for
waterhole.

Figure~\ref{fig:windows-detection_vs_censoring_threshold} shows that structural
filtering improves detection of a phishing attack by $\sim 4x$ (difference
between left and right end points of each curve) over temporal filtering -- by
filtering out alert-FVs from unrelated benign nodes that were active during
the same NTW as infected nodes. Interestingly, the size of a neighborhood
window does not considerably affect the detection when used along with the most
fine-grained structural filtering (treating each recipients' list
individually) -- 3-hour long NTWs results in only a $\sim 12\%$
decrease in the number of compromised nodes (i.e. time to detection). This
shows that there is substantial signal that structural filtering can help
extract from alert-FVs in smaller NTWs (and thus improve Shape GD's time
to detection). 


Structural filtering improves time to detect waterhole attacks as well -- by
5.82x, 4.07x, and 3.75x for 25-, 50-, 100-sec long windows respectively.
Interestingly, structural filtering requires Shape GD to use longer NTWs than
before -- small NTWs (such as 6 seconds from the last sub-section)
no longer supply a sufficient number of alert-FVs for Shape GD to operate
robustly.  Even though structural filtering with a 25 second NTW improves detection
by 5.82x over temporal filtering with 25 second NTWs, the number of infected
nodes at detection time is 139.9 -- higher than the 107 infected nodes for
temporal filtering with a {\em 6 second} NTW (Figure
\ref{fig:waterhole_time_NF}). Temporal and structural filtering thus present
different trade-offs between detection time and work performed by GD -- 
their relative performance is affected by the rate at which true and false positive
FVs are generated.

%
%

%
%

\subsection{Fragility of Count GD}
\label{sec:eval-count}

\begin{figure}[tbp]
\vspace{-0.0in}
   \centering
   \includegraphics[width=0.45\textwidth]{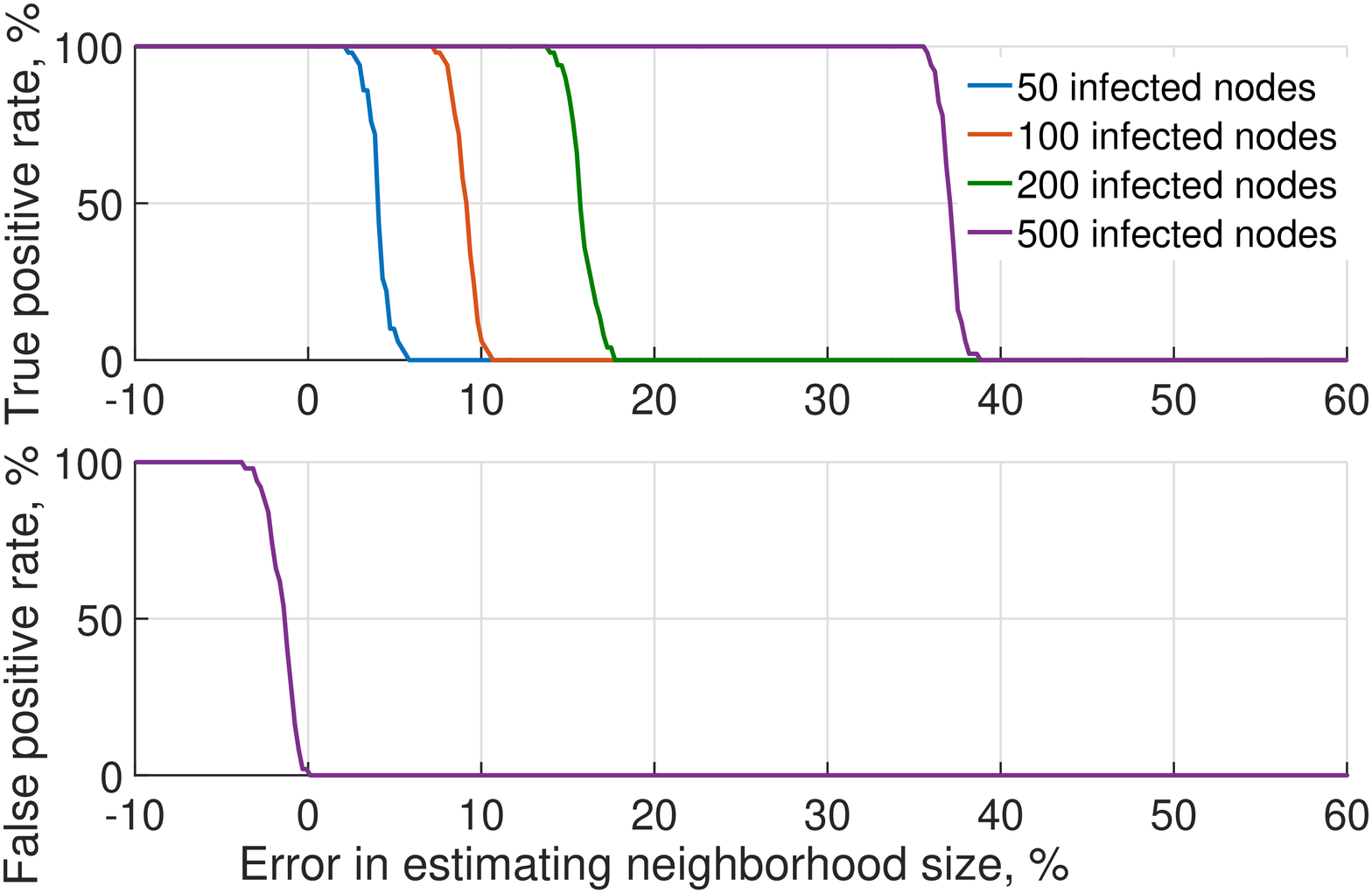}\par
\caption{{\em(Waterhole attack) An error in estimating neighborhood size dramatically affects Count GD's performance. 
It can tolerate at most 0.1\% underestimation errors and 13.8\% overestimation errors to achieve comparable with Shape GD performance.
}}
   \vspace{-0.2in}
\label{fig:windows-noisy_count-waterhole}
\end{figure}

A Count GD algorithm counts the number of alerts over a neighborhood
and compares to a threshold to detect malware. 
This threshold scales linearly in the size of the neighborhood -- 
we now experimentally quantify the error Count GD can tolerate in
phishing (Figure \ref{fig:windows-noisy_count}) and waterhole (Figure
\ref{fig:windows-noisy_count-waterhole}) settings.  Note that the
error in estimating neighborhood size can be double sided --
underestimates (negative error) can make neighborhoods look like alert
hotspots and lead to false positives, while overestimates (positive
error) can lead to missed detections (i.e., lower true positives).

We run Count GD in the same setting as Shape GD when evaluating time-based NF
(Section \ref{sec:time-nf}) -- 30-sec long neighborhood including 17,178 nodes
(Figure \ref{fig:windows-noisy_count-waterhole}) to model a waterhole attack
and 1-hour long neighborhood time window (NTW) with 1086 nodes (Figure
\ref{fig:windows-noisy_count} in Appendix F) to model phishing.  We vary 
infection probability (waterhole)  and click rate in emails (phishing) 
such that the
number of infected nodes in a neighborhood changes from 0 to 55 (phishing) and
from 0 to 500 (waterhole) in four increments -- note that in both scenarios,
only a small fraction (5.5\% and 2.9\%) of nodes per neighborhood get infected
in the worst case.

In this setting, recall that the Shape GD has a maximum global false
positive rate of 1\% and a true positive rate of 100\% -- and
detects malware when only 22 (phishing) and 200 (waterhole) nodes are
infected -- for the same NTWs. When the same number of nodes are
infected, and for a similar detection performance,
our experiments show that the Count GD can only
tolerate neighborhood size estimation errors within a very narrow
range -- [-2\%, 6.3\%] (phishing) and [-0.1\%, 13.8\%]
(waterhole). \ignore{numbers look wrong for phishing -- please fix. Fixed}
A key takeaway here is that underestimating a neighborhood's size
makes Count GD extremely fragile (-2\% in phishing and -0.1\% for
waterhole). On the other hand, overestimating neighborhood sizes
decreases true positives, and this effect is catastrophic by the time
the size estimates err by 17\% (phishing) and 17.5\% (waterhole).

We comment that this effect can be important in practice. Given the
practical deployments where nodes get infected out of band (e.g.,
outside the corporate network), go out of range (with mobile devices),
or with dynamically defined neighborhoods based on actions that can be
missed (e.g. neighborhood defined by nodes that `open' an email instead
of only downloading it from a mail server), the
tight margins on errors can render Count GD extremely unreliable. Even
with sophisticated size estimation algorithms, recall that the
underlying distributions that create these neighborhoods (email open
times, number of clients per server, etc) have sub-exponential heavy
tails \cite{email-resp-dist} -- such distributions typically result in poor parameter
estimates due to lack of higher moments, and thus, poorer statistical
concentrations of estimates about the true value \cite{fkz11}.
Circling back, we see that by eliminating this size dependence compared to
Count GD, our Shape GD provides a robust inference algorithm.

\section{Conclusions}
\label{sec:conclusion}
\vspace{0.1in}
Building robust behavioral detectors is a long-standing problem.  We observe
that attacks on enterprise networks induce a low-dimensional structure on
otherwise high-dimensional feature vectors, but this structure is hard to
exploit because the correlations are hard to predict.  By analyzing alert
feature vectors instead of alerts and filtering the alert-FVs along
neighborhood lines, we amplify the signal buried in correlated feature vectors,
and then use the notion of statistical shape to classify neighborhoods without
having to estimate the expected number of benign and false positive FVs per
neighborhood. We note that both neighborhood-filtering and shape are
complementary techniques that apply across a range of LDs or platforms -- e.g.,
we have determined that \sysname also works well with n-grams based LDs
(instead of histograms) and on the Android platform (in addition to Windows). 

Our methodology composes the traditional host-level malware analysis
methodology with trace-based simulations from real web services (to overcome
the lack of joint LD-GD datasets), and allow us to run sensitivity analyses
that will be precluded by using an actual enterprise trace.  We find that
Shape-GD reduces the number of FPs reported to deeper analyses by
$\sim$100$\times$ and $\sim$200$\times$ when employing time-based filtering
only (for phishing and waterhole scenarios respectively), while structural
filtering reduces alert-FVs to $\sim$1000$\times$ and $\sim$830$\times$
(Appendix~\ref{sec:discussion}).  Neighborhoods and their shape thus serve as a
new and effective lens for dimensionality reduction and significantly improve
false positive rates of state-of-the-art behavioral analyses. 
For example, LDs can operate at a higher false positive rate in order to reduce 
false negatives and improve computation efficiency.


{\footnotesize \bibliographystyle{acm}
\bibliography{global}}

\begin{thebibliography}{10}

\bibitem{phishing-1}
\url{http://www.fraudwatchinternational.com/phishing-alerts}.

\bibitem{cisco-amp}
Advanced malware protection (amp).
\newblock
  \url{http://www.cisco.com/c/en/us/products/security/advanced-malware-protection/index.html}.

\bibitem{dell-ampd}
Advanced malware protection and detection (ampd).
\newblock
  \url{https://www.secureworks.com/capabilities/managed-security/network-security/advanced-malware-protection}.

\bibitem{attack_graphs}
Attack graphs: visualizing 200m alerts a day.
\newblock
  \url{http://on-demand.gputechconf.com/gtc/2016/presentation/s6114-leo-meyerovich-attack-graphs-visualizing-alerts.pdf}.

\bibitem{binning}
Data binning.
\newblock \url{https://en.wikipedia.org/wiki/Histogram}.

\bibitem{enron-dataset}
Enron email dataset.
\newblock \url{https://www.cs.cmu.edu/~./enron}.

\bibitem{yahoo-G4}
G4 - yahoo! network flows data.
\newblock \url{https://webscope.sandbox.yahoo.com}.

\bibitem{graphistry}
Graphistry.
\newblock \url{https://www.iqt.org/graphistry}.

\bibitem{GRR}
Grr rapid response: remote live forensics for incident response.
\newblock \url{https://github.com/google/grr}.

\bibitem{hone_code}
Hone tool.
\newblock \url{https://github.com/pmoody-/Linux-Sensor}.

\bibitem{kaspersky-windows-2015}
Kaspersky security bulletin 2015.
\newblock
  \url{https://securelist.com/files/2015/12/Kaspersky-Security-Bulletin-2015_FINAL_EN.pdf}.

\bibitem{lastline}
Lastline: Defeat advanced malware before it infiltrates your network.
\newblock \url{https://www.lastline.com}.

\bibitem{osquery}
osquery -- performant endpoint visibility.
\newblock \url{https://osquery.io/}.

\bibitem{secure_rank}
Securerank algorithm.
\newblock \url{https://blog.opendns.com/2013/03/28/secure-rank-
  a-large-scale-discovery-algorithm-for-predictive-detection}.

\bibitem{anthem}
Statement regarding cyber attack against anthem.
\newblock
  \url{http://www.sophos.com/en-us/threat-center/mobile-security-threat-report.aspx}.

\bibitem{phishing-2}
Symantec intelligence report.
\newblock
  \url{https://www.symantec.com/content/en/us/enterprise/other_resources/b-intelligence-report-01-2015-en-us.pdf}.

\bibitem{anthem-waterhole}
Symantec report on black vine espionage group.
\newblock
  \url{http://www.symantec.com/content/en/us/enterprise/media/security_response/whitepapers/the-black-vine-cyberespionage-group.pdf}.

\bibitem{virustotal}
Virustotal -- free online virus, malware and url scanner.
\newblock \url{https://www.virustotal.com}.

\bibitem{wasser-wiki}
Wasserstein metric.
\newblock \url{https://en.wikipedia.org/wiki/Wasserstein_metric}.

\bibitem{waterhole_attack}
Why watering hole attacks work.
\newblock
  \url{https://threatpost.com/why-watering-hole-attacks-work-032013/77647}.

\bibitem{kruegel2007}
{\sc Bailey, M., Oberheide, J., Andersen, J., Mao, Z., Jahanian, F., and
  Nazario, J.}
\newblock Automated classification and analysis of internet malware.
\newblock In {\em Recent Advances in Intrusion Detection}. 2007.

\bibitem{benbre00}
{\sc Benamou, J., and Brenier, Y.}
\newblock Mixed l2-wasserstein optimal mapping between prescribed density
  functions.
\newblock {\em Journal of Optimization Theory and Applications\/} (2001).

\bibitem{Bethencourt2005}
{\sc Bethencourt, J., Franklin, J., and Vernon, M.}
\newblock Mapping internet sensors with probe response attacks.
\newblock In {\em Proceedings of the 14th USENIX Security Symposium\/} (2005).

\bibitem{Burguera2011}
{\sc Burguera, I., Zurutuza, U., and Nadjm-Tehrani, S.}
\newblock Crowdroid: Behavior-based malware detection system for android.
\newblock In {\em Proceedings of the 1st ACM Workshop on Security and Privacy
  in Smartphones and Mobile Devices\/} (2011), SPSM.

\bibitem{Canali2012}
{\sc Canali, D., Lanzi, A., Balzarotti, D., Kruegel, C., Christodorescu, M.,
  and Kirda, E.}
\newblock A quantitative study of accuracy in system call-based malware
  detection.
\newblock In {\em Proceedings of the 2012 International Symposium on Software
  Testing and Analysis\/} (2012), ISSTA 2012.

\bibitem{evolutionary_spectral_clustering}
{\sc Chi, Y., Song, X., Zhou, D., Hino, K., and Tseng, B.~L.}
\newblock Evolutionary spectral clustering by incorporating temporal
  smoothness.
\newblock In {\em Proceedings of the 13th ACM SIGKDD International Conference
  on Knowledge Discovery and Data Mining\/} (2007), KDD '07.

\bibitem{Christodorescu:2008:MSM:1342211.1342215}
{\sc Christodorescu, M., Jha, S., and Kruegel, C.}
\newblock Mining specifications of malicious behavior.
\newblock In {\em Proceedings of the 1st India Software Engineering
  Conference\/} (2008), ISEC.

\bibitem{Christodorescu05}
{\sc Christodorescu, M., Jha, S., Seshia, S.~A., Song, D., and Bryant, R.~E.}
\newblock Semantics-aware malware detection.
\newblock In {\em Proceedings of the 2005 IEEE Symposium on Security and
  Privacy\/} (2005).

\bibitem{clark2013}
{\sc Clark, S.~S., Ransford, B., Rahmati, A., Guineau, S., Sorber, J., Xu, W.,
  and Fu, K.}
\newblock {W}atts{U}p{D}oc: Power side channels to nonintrusively discover
  untargeted malware on embedded medical devices.
\newblock In {\em USENIX Workshop on Health Information Technologies\/} (2013).

\bibitem{Dash2006}
{\sc Dash, D., Kveton, B., Agosta, J.~M., Schooler, E., Chandrashekar, J.,
  Bachrach, A., and Newman, A.}
\newblock When gossip is good: Distributed probabilistic inference for
  detection of slow network intrusions.
\newblock In {\em Proceedings of the 21st National Conference on Artificial
  Intelligence\/} (2006).

\bibitem{Demme2013}
{\sc Demme, J., Maycock, M., Schmitz, J., Tang, A., Waksman, A., Sethumadhavan,
  S., and Stolfo, S.}
\newblock On the feasibility of online malware detection with performance
  counters.
\newblock In {\em Proceedings of the 40th Annual International Symposium on
  Computer Architecture\/} (2013).

\bibitem{donoho1983notion}
{\sc Donoho, D.~L., and Huber, P.~J.}
\newblock The notion of breakdown point.
\newblock {\em A festschrift for Erich L. Lehmann 157184\/} (1983).

\bibitem{hone}
{\sc Fink, G.~A., Duggirala, V., Correa, R., and North, C.}
\newblock Bridging the host-network divide: Survey, taxonomy, and solution.
\newblock In {\em Proceedings of the 20th Conference on Large Installation
  System Administration\/} (Berkeley, CA, USA, 2006), LISA '06, USENIX
  Association, pp.~20--20.

\bibitem{Forrest1996}
{\sc Forrest, S., Hofmeyr, S., Somayaji, A., and Longstaff, T.}
\newblock A sense of self for unix processes.
\newblock In {\em Security and Privacy, 1996. Proceedings., IEEE Symposium
  on\/} (1996).

\bibitem{fkz11}
{\sc Foss, S., Korshunov, D., and Zachary, S.}
\newblock An introduction to heavy-tailed and subexponential distributions,
  2009.
\newblock Springer Series in Operations Research and Financial Engineering.

\bibitem{Fredrikson2010}
{\sc Fredrikson, M., Jha, S., Christodorescu, M., Sailer, R., and Yan, X.}
\newblock Synthesizing near-optimal malware specifications from suspicious
  behaviors.
\newblock In {\em IEEE Symposium on Security and Privacy\/} (2010).

\bibitem{wenkee_lee_bothunter_2007}
{\sc Gu, G., Porras, P., Yegneswaran, V., Fong, M., and Lee, W.}
\newblock Bothunter: Detecting malware infection through ids-driven dialog
  correlation.
\newblock In {\em Proceedings of 16th USENIX Security Symposium\/} (2007).

\bibitem{BotSniffer}
{\sc Gu, G., Zhang, J., and Lee, W.}
\newblock Botsniffer: Detecting botnet command and control channels in network
  traffic.
\newblock In {\em Presented at the 16th Annual Network \& Distributed System
  Security Symposium\/} (2008), NDSS.

\bibitem{network_intrusion_detection}
{\sc Handley, M., Paxson, V., and Kreibich, C.}
\newblock Network intrusion detection: Evasion, traffic normalization, and
  end-to-end protocol semantics.
\newblock In {\em Proceedings of the 10th Conference on USENIX Security
  Symposium - Volume 10\/} (Berkeley, CA, USA, 2001), SSYM'01, USENIX
  Association.

\bibitem{Juxtapp}
{\sc Hanna, S., Huang, L., Wu, E., Li, S., Chen, C., and Song, D.}
\newblock Juxtapp: A scalable system for detecting code reuse among android
  applications.
\newblock In {\em Detection of Intrusions and Malware, and Vulnerability
  Assessment}. 2013.

\bibitem{mutant_x}
{\sc Hu, X., Bhatkar, S., Griffin, K., and Shin, K.~G.}
\newblock Mutantx-s: Scalable malware clustering based on static features.
\newblock In {\em Proceedings of the 2013 USENIX Conference on Annual Technical
  Conference\/} (Berkeley, CA, USA, 2013), USENIX ATC'13, USENIX Association,
  pp.~187--198.

\bibitem{huber2011robust}
{\sc Huber, P.~J.}
\newblock {\em Robust statistics}.
\newblock Springer, 2011.

\bibitem{kaufman_clustering}
{\sc Kaufman, L., and Rousseeuw, P.~J.}
\newblock {\em Finding Groups in Data: An Introduction to Cluster Analysis.}
\newblock John Wiley, 1990.

\bibitem{dmitry-ensemble}
{\sc Khasawneh, K.~N., Ozsoy, M., Donovick, C., Abu-Ghazaleh, N., and
  Ponomarev, D.}
\newblock Ensemble learning for low-level hardware-supported malware detection.
\newblock In {\em Research in Attacks, Intrusions, and Defenses}. Springer
  International Publishing, 2015, pp.~3--25.

\bibitem{kruegel-bare-metal}
{\sc Kirat, D., Vigna, G., and Kruegel, C.}
\newblock Barecloud: Bare-metal analysis-based evasive malware detection.
\newblock In {\em Proceedings of the 23rd USENIX Conference on Security
  Symposium\/} (2014).

\bibitem{Kolbitsch:2009:EEM:1855768.1855790}
{\sc Kolbitsch, C., Comparetti, P.~M., Kruegel, C., Kirda, E., Zhou, X., and
  Wang, X.}
\newblock Effective and efficient malware detection at the end host.
\newblock In {\em Proceedings of the 18th Conference on USENIX Security
  Symposium\/} (2009).

\bibitem{email-resp-dist}
{\sc Kooti, F., Aiello, L.~M., Grbovic, M., Lerman, K., and Mantrach, A.}
\newblock Evolution of conversations in the age of email overload.
\newblock In {\em Proceedings of 24th International World Wide Web Conferenced
  (WWW)\/} (2015).

\bibitem{krugel2001}
{\sc Kr\"{u}egel, C., Toth, T., and Kerer, C.}
\newblock Decentralized event correlation for intrusion detection.
\newblock In {\em Proceedings of the 4th International Conference Seoul on
  Information Security and Cryptology\/} (2002), ICISC '01, Springer-Verlag.

\bibitem{downloader_graphs}
{\sc Kwon, B.~J., Mondal, J., Jang, J., Bilge, L., and Dumitras, T.}
\newblock The dropper effect: Insights into malware distribution with
  downloader graph analytics.
\newblock In {\em Proceedings of the 22Nd ACM SIGSAC Conference on Computer and
  Communications Security\/} (New York, NY, USA, 2015), CCS '15, ACM,
  pp.~1118--1129.

\bibitem{Locasto2005}
{\sc Locasto, M., Parekh, J., Keromytis, A., and Stolfo, S.}
\newblock Towards collaborative security and p2p intrusion detection.
\newblock In {\em Information Assurance Workshop, IAW\/} (2005).

\bibitem{Christodorescu06}
{\sc Mihai~Christodorescu, S.~J.}
\newblock Static analysis of executables to detect malicious patterns.
\newblock Tech. rep., The University of Wisconsin, Madison, 2006.

\bibitem{vt_report_classification}
{\sc Miller, B., Kantchelian, A., Tschantz, M.~C., Afroz, S., Bachwani, R.,
  Faizullabhoy, R., Huang, L., Shankar, V., Wu, T., Yiu, G., Joseph, A.~D., and
  Tygar, J.~D.}
\newblock Reviewer integration and performance measurement for malware
  detection.
\newblock In {\em Proceedings of the 13th International Conference on Detection
  of Intrusions and Malware, and Vulnerability Assessment - Volume 9721\/} (New
  York, NY, USA, 2016), DIMVA 2016, Springer-Verlag New York, Inc.,
  pp.~122--141.

\bibitem{bot_grep}
{\sc Nagaraja, S., Mittal, P., Hong, C.-Y., Caesar, M., and Borisov, N.}
\newblock Botgrep: Finding p2p bots with structured graph analysis.
\newblock In {\em Proceedings of the 19th USENIX Conference on Security\/}
  (Berkeley, CA, USA, 2010), USENIX Security'10, USENIX Association, pp.~7--7.

\bibitem{on_spectral_clustering_ng_01}
{\sc Ng, A.~Y., Jordan, M.~I., and Weiss, Y.}
\newblock On spectral clustering: Analysis and an algorithm.
\newblock In {\em ADVANCES IN NEURAL INFORMATION PROCESSING SYSTEMS\/} (2001).

\bibitem{mining_log_data_alina_oprea}
{\sc Oprea, A., Li, Z., Yen, T.-F., Chin, S.~H., and Alrwais, S.}
\newblock Detection of early-stage enterprise infection by mining large-scale
  log data.
\newblock In {\em Proceedings of the 2015 45th Annual IEEE/IFIP International
  Conference on Dependable Systems and Networks\/} (Washington, DC, USA, 2015),
  DSN '15, IEEE Computer Society, pp.~45--56.

\bibitem{practical_black_box_attacks}
{\sc Papernot, N., McDaniel, P., Goodfellow, I., Jha, S., Celik, Z.~B., and
  Swami, A.}
\newblock Practical black-box attacks against machine learning.
\newblock In {\em Proceedings of the 2017 ACM on Asia Conference on Computer
  and Communications Security\/} (New York, NY, USA, 2017), ASIA CCS '17, ACM,
  pp.~506--519.

\bibitem{paxson1999}
{\sc Paxson, V.}
\newblock Bro: A system for detecting network intruders in real-time.
\newblock {\em Comput. Netw. 31}, 23-24 (1999), 2435--2463.

\bibitem{malware_distribution_net}
{\sc Provos, N., Mavrommatis, P., Rajab, M.~A., and Monrose, F.}
\newblock All your iframes point to us.
\newblock In {\em Proceedings of the 17th Conference on Security Symposium\/}
  (Berkeley, CA, USA, 2008), SS'08, USENIX Association, pp.~1--15.

\bibitem{robertson_anomaly}
{\sc Robertson, W., Maggi, F., Kruegel, C., and Vigna, G.}
\newblock Effective anomaly detection with scarce training data.
\newblock In {\em Proceedings of the Network and Distributed System Security
  Symposium (NDSS)\/} (2010).

\bibitem{Shin_Infocom12_EFFORT}
{\sc Shin, S., Xu, Z., and Gu, G.}
\newblock {EFFORT: Efficient and Effective Bot Malware Detection}.
\newblock In {\em Proceedings of the 31th Annual IEEE Conference on Computer
  Communications (INFOCOM'12) Mini-Conference\/} (March 2012).

\bibitem{Shinoda2005}
{\sc Shinoda, Y., Ikai, K., and Itoh, M.}
\newblock Vulnerabilities of passive internet threat monitors.
\newblock In {\em Proceedings of the 14th Conference on USENIX Security
  Symposium\/} (2005).

\bibitem{Shmatikov2007}
{\sc Shmatikov, V., and Wang, M.-H.}
\newblock Security against probe-response attacks in collaborative intrusion
  detection.
\newblock In {\em the Workshop on Large Scale Attack Defense\/} (2007).

\bibitem{Sommer2010}
{\sc Sommer, R., and Paxson, V.}
\newblock Outside the closed world: On using machine learning for network
  intrusion detection.
\newblock In {\em the IEEE Symposium on Security and Privacy\/} (2010).

\bibitem{evade-pdf}
{\sc {\v{S}}rndi{\'c}, N., and Laskov, P.}
\newblock Practical evasion of a learning-based classifier: A case study.
\newblock In {\em the IEEE Symposium on Security and Privacy\/} (2014).

\bibitem{Tang2014}
{\sc Tang, A., Sethumadhavan, S., and Stolfo, S.}
\newblock Unsupervised anomaly-based malware detection using hardware features.
\newblock In {\em Research in Attacks, Intrusions and Defenses}. 2014.

\bibitem{vallender1974calculation}
{\sc Vallender, S.}
\newblock Calculation of the wasserstein distance between probability
  distributions on the line.
\newblock {\em Theory of Probability \& Its Applications 18}, 4 (1974),
  784--786.

\bibitem{vlachos2004}
{\sc Vlachos, V., Androutsellis-Theotokis, S., and Spinellis, D.}
\newblock Security applications of peer-to-peer networks.
\newblock {\em Comput. Netw. 45}, 2 (2004).

\bibitem{spectral_clustering_tutorial_2007}
{\sc von Luxburg, U.}
\newblock A tutorial on spectral clustering.
\newblock {\em Statistics and Computing\/} (2007).

\bibitem{Wagner2002}
{\sc Wagner, D., and Soto, P.}
\newblock Mimicry attacks on host-based intrusion detection systems.
\newblock In {\em the ACM Conference on Computer and Communications Security\/}
  (2002).

\bibitem{Seurat2004}
{\sc Xie, Y., Kim, H., O'Hallaron, D.~R., Reiter, M.~K., and Zhang, H.}
\newblock Seurat: {A} pointillist approach to anomaly detection.
\newblock In {\em Recent Advances in Intrusion Detection\/} (2004).

\bibitem{xu2013outlier}
{\sc Xu, H., Caramanis, C., and Mannor, S.}
\newblock Outlier-robust pca: the high-dimensional case.
\newblock {\em IEEE transactions on information theory 59}, 1 (2013), 546--572.

\bibitem{evade-ml}
{\sc Xu, W., Qi, Y., and Evans, D.}
\newblock Automatically evading classifiers: A case study on pdf malware
  classifiers.
\newblock In {\em Network and Distributed Systems Symposium\/} (2016).

\bibitem{beehive2013}
{\sc Yen, T.-F., Oprea, A., Onarlioglu, K., Leetham, T., Robertson, W., Juels,
  A., and Kirda, E.}
\newblock Beehive: Large-scale log analysis for detecting suspicious activity
  in enterprise networks.
\newblock In {\em Proceedings of the 29th Annual Computer Security Applications
  Conference\/} (2013), ACSAC '13, ACM, pp.~199--208.

\bibitem{zhang2001}
{\sc Zhang, Z., Li, J., Manikopoulos, C.~N., Jorgenson, J., and Ucles, J.}
\newblock Hide: a hierarchical network intrusion detection system using
  statistical preprocessing and neural network classification.
\newblock In {\em the IEEE Workshop on Information Assurance and Security\/}
  (2001).

\bibitem{bot_graph}
{\sc Zhao, Y., Xie, Y., Yu, F., Ke, Q., Yu, Y., Chen, Y., and Gillum, E.}
\newblock Botgraph: Large scale spamming botnet detection.
\newblock In {\em Proceedings of the 6th USENIX Symposium on Networked Systems
  Design and Implementation\/} (Berkeley, CA, USA, 2009), NSDI'09, USENIX
  Association, pp.~321--334.

\end{thebibliography}
\appendix
\section{How accurate is clustering for global malware detection?}
\label{sec:eval-clustering}
\ignore{
\begin{figure}[t]
\includegraphics[width=0.48\textwidth]{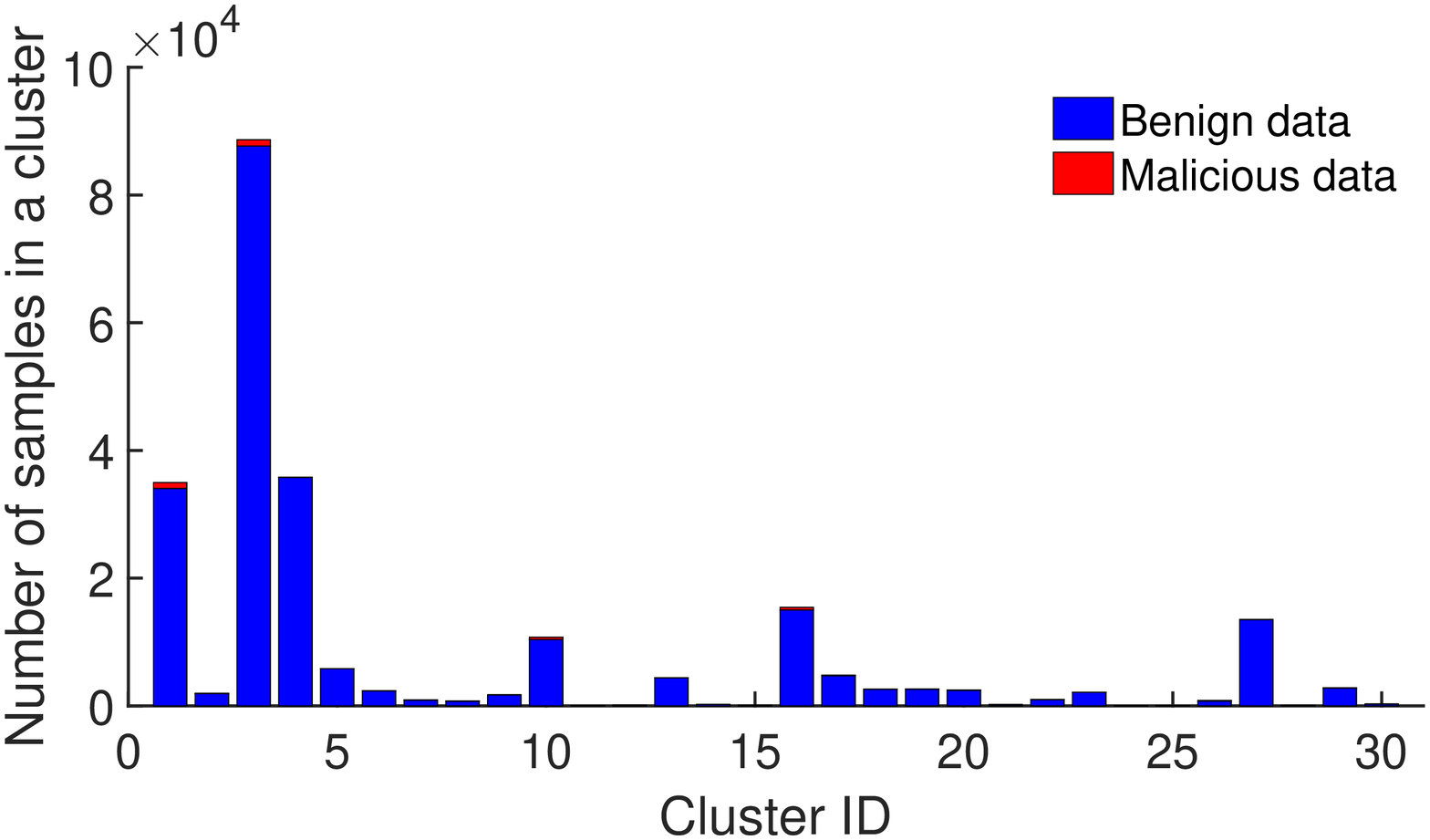}\par
\caption{(Waterhole attack) A common clustering approach \cite{beehive2013, Seurat2004, kaufman_clustering} fails to
  distinguish between malicious and benign feature vectors in our
  setting, i.e. each cluster includes significant number of both
  malicious and benign FVs.}
  \vspace{-0.2in}  
\label{fig:clustering-bars-waterhole}
\end{figure}
}


While Count GD is fragile, clustering GDs are inaccurate in the early stages of
infection.  This is why prior work~\cite{beehive2013} uses clustering to
(offline) identify high-priority incidents from security logs for human
analysis (instead of as an always-on GD) -- this use case is complementary to
an always-on global detector. 
We quantify a recent clustering GD's~\cite{beehive2013} detection rate on our
data set. 


First, we reduce dimensionality of 390-dimensional FVs by projecting them on
the top 10 PCA components, which retain 95.72\% of the data variance.  Second,
we use an adaptation of the K-means clustering algorithm that does not require
specifying the number of clusters in advance \cite{beehive2013, Seurat2004,
kaufman_clustering}. Specifically, the algorithm consists of the following
three steps: (1) select a vector at random as the first centroid and assign all
vectors to this cluster; (2) find a vector furthest away from its centroid 
(following Beehive~\cite{beehive2013}, we use L1 distance) and
make it a center of a new cluster, and reassign every vector to the cluster
with the closest centroid; and (3) repeat step 2 until no vector is further
away from its centroid than half of the average inter-cluster distance.


The evaluation settings of the clustering algorithm match exactly the settings
where Shape GD detects infected neighborhoods with 99\% confidence.
Specifically, the algorithm clusters the data that we collected in a
17,178-node neighborhood under a waterhole attack within 30 seconds and
the data that we collected over an hour-long session across 1086 nodes in a
medium size corporate network under a phishing attack
(Section~\ref{sec:datasets}).  As we have already demonstrated
(Section~\ref{sec:time-nf}), Shape GD starts detecting malware when 107
(waterhole attack) and 22 (phishing attack) nodes get compromised (as in
experiments for Figures~\ref{fig:windows-shape_vs_count}
and~\ref{fig:windows-shape_vs_count-waterhole}).

\begin{figure}[t]
\includegraphics[width=0.48\textwidth]{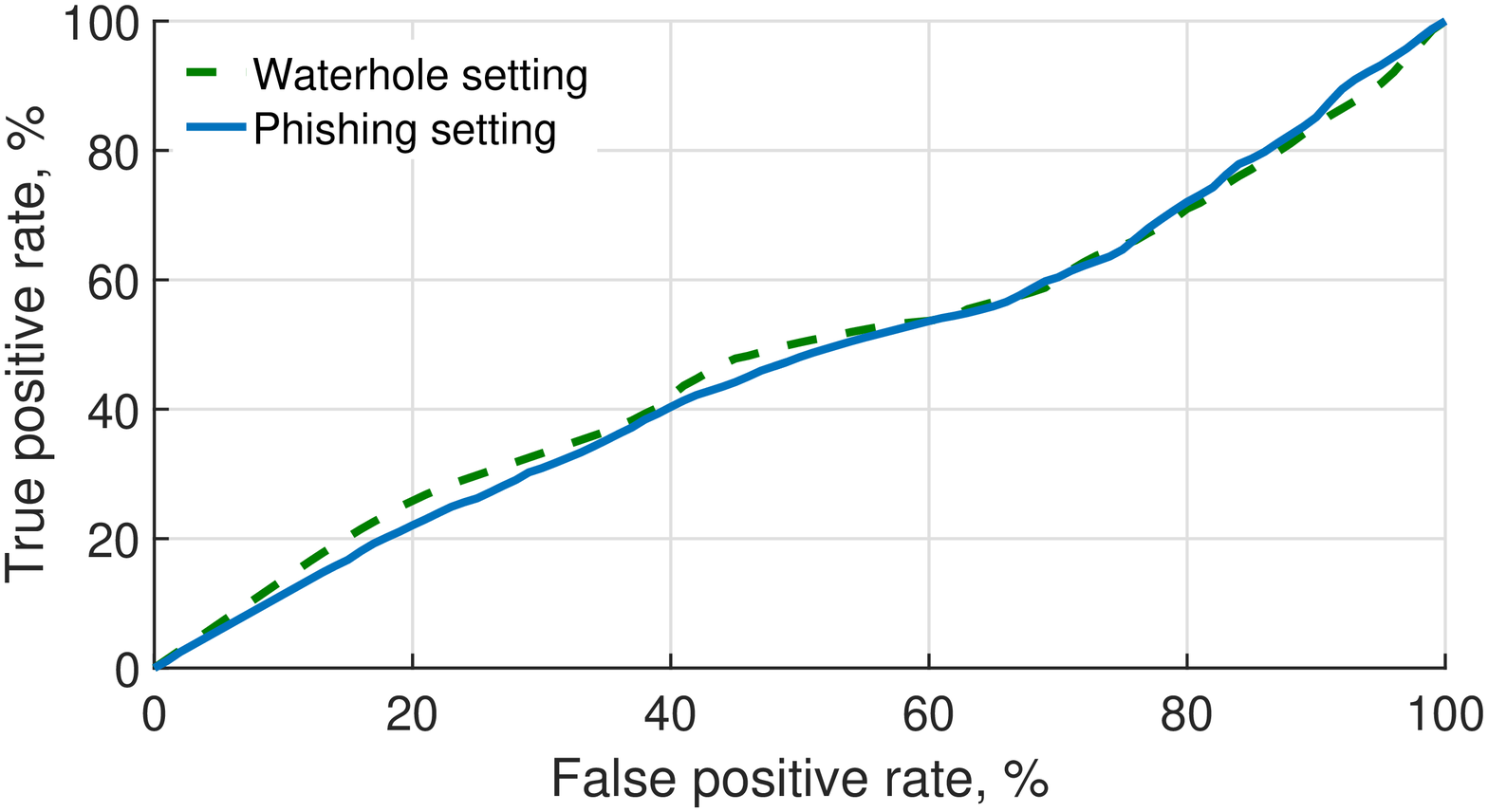}\par
\caption{(ROC curve) True positive v. False positive curve shows detection
accuracy of the clustering-based malware detector
\cite{beehive2013}. Its Area Under the Curve (AUC) parameter averaged for 10 runs reaches
only 48.3\% and 47.4\% in the case of waterhole and phishing attacks respectively; such low AUC value makes it unusable as a
global detector.} 
\vspace{-0.3in}
\label{fig:clustering-roc}
\end{figure}

Clustering does not fare well, and 
results look very similar for both waterhole and phishing
experiments.
The clustering algorithm partitions waterhole data set into 30 clusters.  We
observe three large clusters that aggregate most of the benign FVs. However,
the algorithm fails to find small 'outlying' clusters consisting of
predominantly malicious data.  As for the phishing experiment, we observe a
similar picture: the algorithm forms slightly higher number of clusters -- 33
rather than 30 -- and it identifies 4 densely populated clusters.  In both
cases each cluster heavily mixes benign and malicious data, hence the
clustering approach suffers from poor discriminative ability, i.e. it is unable
to separate malicious and benign samples.

Note the clustering algorithm enforces explicit ordering across the clusters.
That is, the algorithm forms a new cluster around an FV that is furthest away
from its cluster centroid. Thus, earlier a cluster is created, the more
suspicious it is.  By design of the clustering algorithm, the clusters are
subject to a deeper analysis in order of their suspiciousness.  
Such an inherent ordering allows us to build a receiver operating curve (Figure
\ref{fig:clustering-roc}) and compute a typical metric for a binary classifier
-- Area Under the Curve (AUC) by averaging across 10 runs.  
The AUC reaches only 48.3\% and 47.4\% for waterhole and phishing experiments
respectively.  

This experiment illustrates the failure of the traditional recipe of
dimensionality reduction plus clustering. There is a fundamental reason for
this -- the neighborhoods
we seek to detect are small compared to the total number of nodes in the
system.  
Optimization-based algorithms that exploit density, including
K-Means and related algorithms, fail to detect small clusters in high
dimensions, even under dimensionality reduction.  The reason is that the
dimensionality reduction is either explicitly random (e.g., as in
Johnson-Lindenstrauss type approaches), or, if data-dependent (like PCA), it is
effectively independent of small clusters, as these represent very little of
the energy (the variance) of the overall data. Spectral clustering style
algorithms~\cite{spectral_clustering_tutorial_2007,on_spectral_clustering_ng_01,evolutionary_spectral_clustering}
are also notoriously unable to deal with highly
unbalanced sized clusters, and in particular, are unable to find small
clusters.

Shape GD also reduces dimensionality but does so after neighborhood
filtering.  This amplifies the impact of small neighborhoods. The combination
of dimensionality reduction, small-neighborhood-amplification, and then
aggregation represents a novel approach to this detection problem, and our
experiments validate this intuition.



\section{Local Detectors}
\label{sec:LDs}

Our first step is to establish a good local detector (LD) for desktop systems
running Windows OS.  In particular, we choose system call based LDs since the
system call interface has visibility into an app's intercation with core OS
components -- file system, Windows registry, network -- and
can thus capture signals relevant to malware executions.

We experiment with an extensive set of system-call LDs -- our takeaway is that
even the best LD we could construct operates at a true- and false-positive
ratio of 92.4\%:6\% and is not deployable by itself (i.e.,
will create $\sim$30 false positives every 10 minutes without a GD). 



Each LD comprises of a feature extraction (FE) algorithm and a machine learning
(ML) classifier.  Our FE algorithm partitions the time-series of system calls
into 1-sec chunks and represents each chunk as a histogram (where each bin
contains frequency of a particular system call).  Then it projects all feature
vectors onto 10-dimensional subspace spanned by top 10 principal components
generated by PCA algorithm.  We choose ML
classifiers (used throughout prior work 
because these are computationally efficient to train) such as SVMs, random
forest, k-Nearest Neighbors, etc, and do not include complex alternatives such
as artificial neural networks or deep learning algorithms.  We also
deliberately avoid handcrafted ML algorithms and hardcoded detection rules.



\begin{figure}[tbp]
   \centering
   \includegraphics[width=0.45\textwidth]{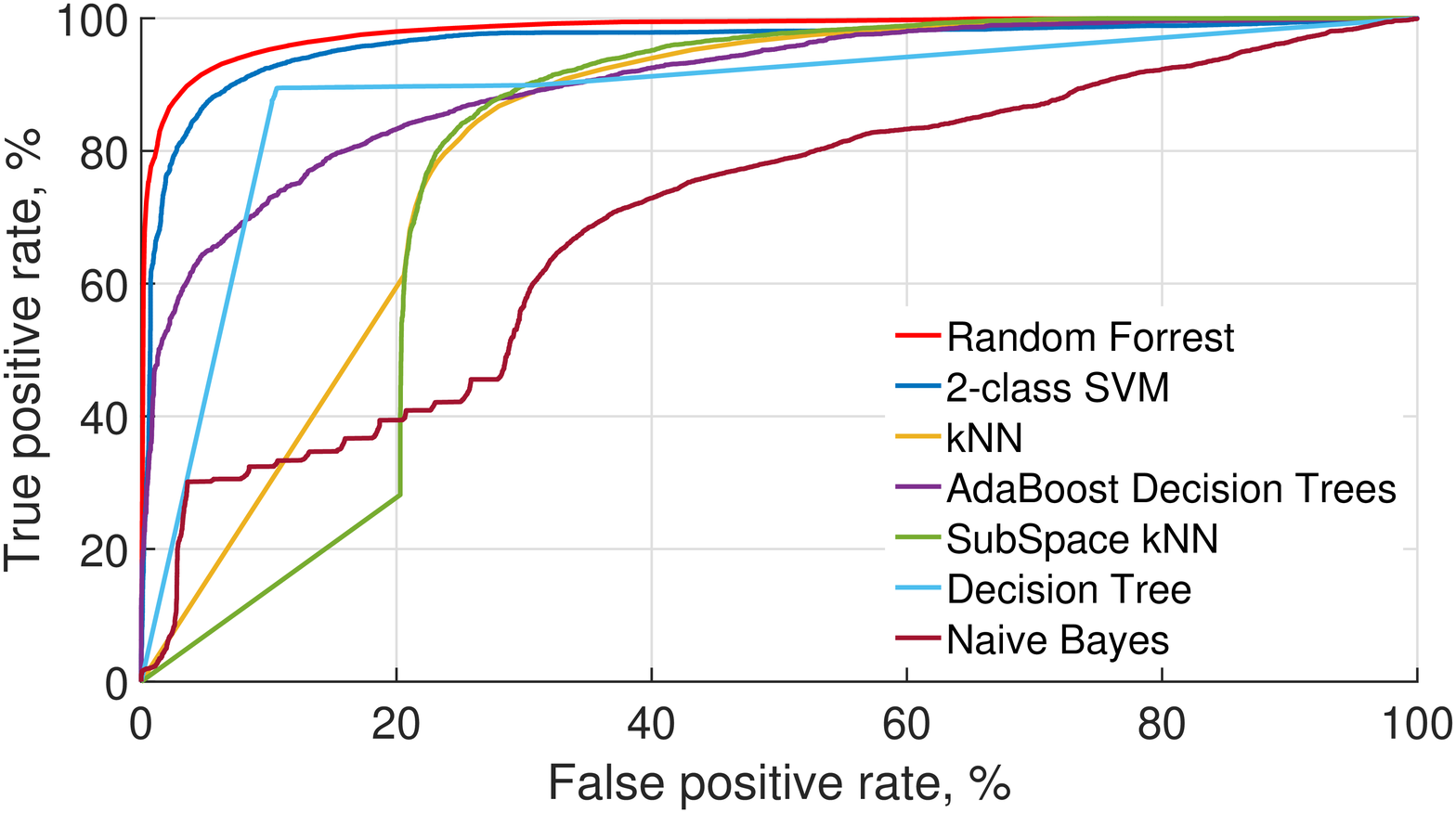}\par
\caption{(ROC curves) True positive v. False positive curves shows detection
accuracy of seven local detectors.  
Random Forest outperforms all others;
but has unacceptably high false
positive rate (above 10\%) if one wants to achieve at least 95\% true positive
rate.}
   \vspace{-0.2in}
\label{fig:windows-LD-roc}
\end{figure}

Figure~\ref{fig:windows-LD-roc} plots the true positive v. false positive rates
(i.e. the ROC curves) of the seven ML algorithms we evaluate.  The area under
the ROC curve (AUC) is a quantitative measure of LD's performance: the larger
the AUC, the more accurate the detector.  We specifically experiment with seven
state-of-the-art ML algorithms: random forest, 2-class SVM, kNN, decision
trees, naive Bayes, and their ensemble versions -- boosted decision trees with
AdaBoost algorithm and Random SubSpace ensemble of kNN classifiers
(Figure~\ref{fig:windows-LD-roc}).  We also evaluated 1-class SVM as an anomaly
detector -- however, it yielded an extremely high FP rate and we exclude it
from further discussion.  Overall, the random forest classifier worked best --
it has the highest AUC and we pick an operating point of
92.4\% true positives at a false positive rate of 6\%. 

\section{How many FVs does Shape GD need to make robust predictions?} 
\label{sec:nbd-size}



The number of FVs per neighborhood required by \sysname to make a robust
prediction is a crucial parameter.  With too few FVs produced by a
neighborhood, benign neighborhoods' ShapeScore will have high variance (i.e.
benign distribution in Figure~\ref{fig:windows-hist} becomes wide and the gap
between two distributions shrinks), leading to global false positives and
negatives.
On the other hand, if neighborhoods are large, their ShapeScores
will be dominated by the large number of benign FVs and thus lead to
missed alerts (false negatives) especially in the early stages of
infection. 
Further, the number of alert-FVs generated by a neighborhood in a
deployed \sysname need to be comparable to or larger than those used
in training -- hence, we want to determine the smallest
number of FVs \sysname needs to make a robust prediction.

\begin{figure}[tbp]
   \centering
   \includegraphics[width=0.45\textwidth]{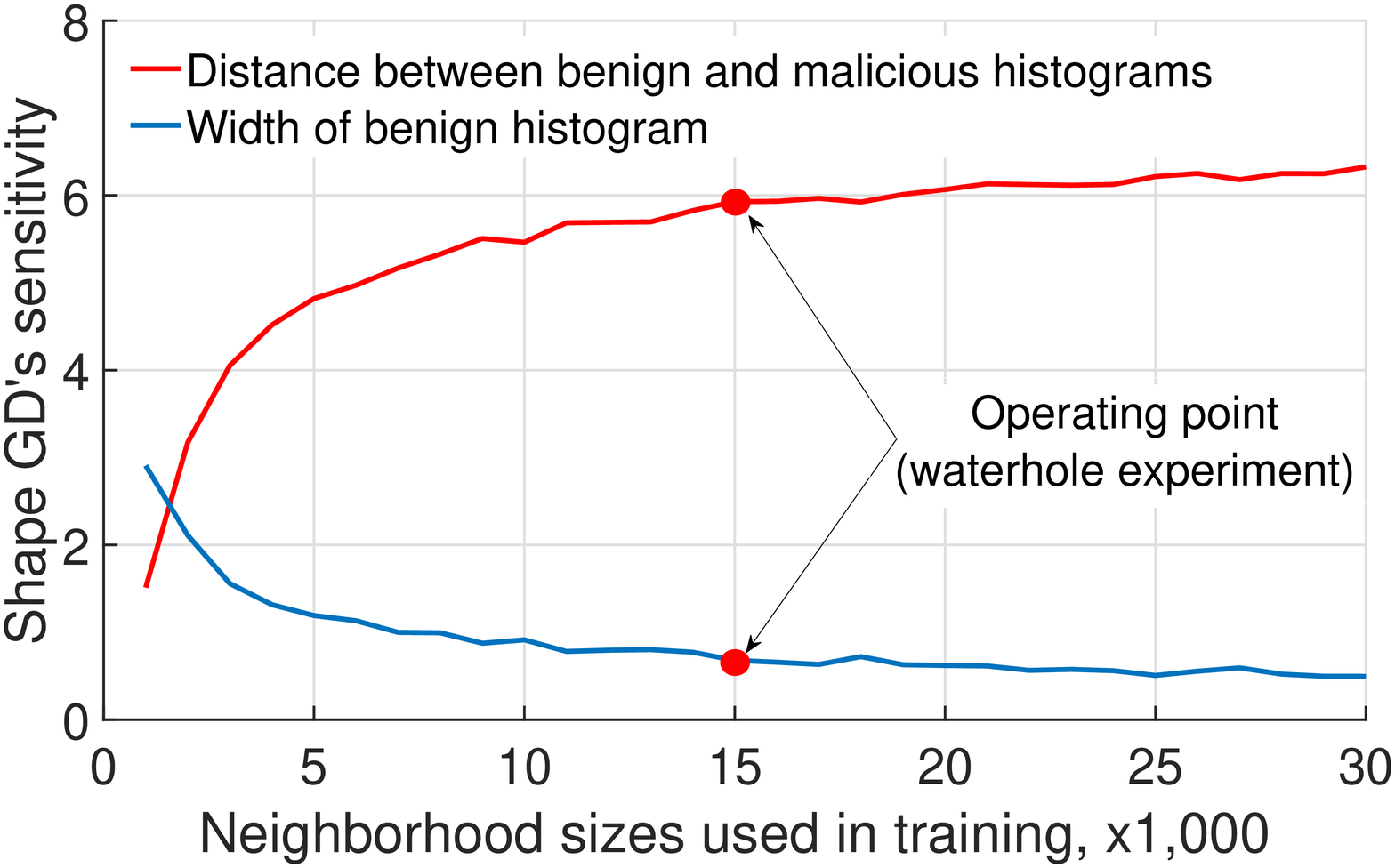}\par
   \caption{Analysis of ShapeScore histogram parameters when changing
     neighborhood size.  The curves flatten out on the right side from
     the operating point.}
   \vspace{-0.2in}
\label{fig:windows-hist-stability}
\end{figure}

Figure~\ref{fig:windows-hist-stability} shows the sensitivity of \sysname to
neighborhood size (i.e., the number of FVs generated by nodes in a neighborhood
during training stage).  We 
vary the number of FVs that a neighborhood generates from 3,000 up to
30,000 FVs and average the results of 10 experiments.
We present two metrics in Figure~\ref{fig:windows-hist-stability} -- the red
curve plots the inter-class distance (between histograms of benign and
malicious neighborhoods from Figure~\ref{fig:windows-hist}), and the blue curve
plots intra-class distance (i.e. the width of the benign histogram).
%
%
Figure~\ref{fig:windows-hist-stability} shows that the red inter-class distance
increases (and blue intra-class variance decreases) quickly as neighborhood
size increases, and both curves
flatten out once the neighborhoods start generating more than 15,000
FVs.


This shows that (for our Windows programs dataset) neighborhoods generating
15,000 FVs or more are a good choice to train Shape GD because purely malicious
or benign distributions stabilize at this size.  In real scenarios with
mixtures of mostly benign and a few malicious neighborhoods, the number of FVs
will have to be scaled up depending upon the timescale of attacks (hours for
phishing v. seconds for waterhole) and the number of nodes affected by an
attack (tens of nodes  in enterprise email networks v.  thousands in a broader
waterhole attack on the enterprise). In the phishing and waterhole attack case
studies in the paper, we use neighborhoods of 1,086 and $\sim$17,000 nodes that
produce 15k FVs and 100k FVs respectively.





\section{Computation and Communication Costs of Shape-GD}
\label{sec:overhead-appendix}
\textbf{Local detectors.} Generating a single FV, which is a 1-sec
histogram of system calls, on a local host is equivalent to performing
2,500 (system call frequency) direct table lookups on average and
incrementing corresponding counters. Projection on a PCA basis
requires computing 10 dot products. Finally, running an LD, which is
Random Forest in our case, results in performing 330 scalar
comparisons on average. At 1 second per FV, the overheads of such an
LD are negligible.

\textbf{Data transfer.} Each FV is composed of 10 floating point
numbers (40 bytes total if assuming single precision format). In the
phishing experiment 1086 hosts transfer (in aggregate) $\sim 40
KB/sec$; data transfer rate in waterhole setting is a little bit
higher: $\sim 4,450$ hosts transfer (in aggregate) $\sim 174 KB/sec$.
In both cases we assume Shape GD using pure time-based filtering with
1 hour and 6 sec neighborhood time windows respectively.

If Shape GD employs structural filtering on top of the time-based one,
then data transfer depends on the number of emails floating in a
network or on the number of servers. In both cases, data transfer
scales linearly with the number of emails and servers. When applying
the most fine-grained structural filtering in our experiments, the
nodes susceptible to phishing attacks transfer $\sim 4 KB/sec$ per
email and the nodes susceptible to waterhole attacks send $\sim 40
KB/sec$ per server when using 1 hour and 25 sec neighborhood windows
respectively.

\textbf{Server computations.} After receiving a batch of alert-FVs,
Shape GD performs lightweight computations.
Overhead of binning scales linearly with the number of alert-FVs in a
batch; each binning operation is a direct table lookup together with
counter increment. Calculating ShapeScore, which is Wasserstein
distance, results in a sequence of addition operations, whose total
number is equal to the dimensionality of FVs, which is 10, multiplied
by the number of bins, which is 50. To summarize, Shape GD's computational
requirements are fairly light-weight.

\section{Discussion}
\label{sec:discussion}

\noindent \textbf{Global FPs vs LD FPs.} As remarked in the Introduction, an FP
of 1\% at the global level means that we will see one alert every 100 - 300
hours (for the phishing scenario) and 100 seconds (for waterhole scenario 
the neighborhood time window slides by 1 second). This reduces work to 
be performed by the deeper, second-level analysis considerably.

Specifically, LDs operating at 6\% false positive rate generate 23.5M -- 70M
FPs within 100--300 hours time interval in a network of 1086 nodes (phishing)
and 300K alerts within every 100 sec interval where neighborhoods include
$\sim$50K nodes on average (waterhole). Shape GD filters these alerts.  When
using 1--3 hours (phishing) and 6 sec (waterhole) time-based neighborhood
filtering, Shape GD will report to a system running a deeper analysis
approximately 234.5K -- 703.7K FPs raised by LDs (phishing) and approximately
1.4K FPs (waterhole).  Adding structural filtering brings these numbers down to
21.6K -- 64.8K FPs (phishing) and 360 FPs (waterhole).

Compared to a neighborhood of LDs, Shape GD thus reduces  the number of FPs
reported to deeper analyses by $\sim$100$\times$ and $\sim$200$\times$ when
employing time-based filtering only (for phishing and waterhole scenarios
respectively), while structural filtering reduces alert-FVs for deeper analysis
to $\sim$1000$\times$ and $\sim$830$\times$.
In both scenarios, analysts can choose to reduce
number of alert-FVs to be analyzed by sliding neighborhood windows by a larger
interval; however, this will increase the time to detect malware infection.

\noindent \textbf{Shape property across LDs and platforms.} \sysname relies on conditional
separability of FPs and TPs, and we use only one LD type for evaluation -- a
system call histogram-based LD.  However, we have experimentally determined
that FPs and TPs are separable for other LD types as well -- an n-gram-based
LD~\cite{mutant_x} and an LD that uses VirusTotal~\cite{virustotal} reports for
malware detection~\cite{vt_report_classification}.  Further, we can classify
malicious neighborhoods on the Android platform -- using malware binaries
obtained from the NCSU dataset and contagio dump website, and using benign
applications that we drive using real human user input -- in addition to the
Windows setup that we describe here. We have left out the details due to lack
of space but can produce an anonymous report if requested.

Though the local detectors we built have a 6\% FP rate, \sysname can work well
work with better LDs. \sysname only requires LD's FPs and TPs to be separable
and to be able to aggregate enough alert-FVs across the nodes within a
neighborhood.  We deliberately do not consider rule-based LDs that are commonly
used within enterprise networks because, even though their FP rate is very low,
they suffer from a high false negative rate, and they can be easily evaded with
simple malware transformations.

\noindent{\bf Performance Overheads.} Recall that Shape GD requires only 
alert FVs -- this leads to a two-fold dimensionality reduction
when sending data from individual LDs to the GD. First, the FVs are
low-dimensional (here, 10-dimensional vectors).  Second, only alert FVs are
needed -- this leads to a 16-fold reduction in data (roughly only 6\% of the
FVs lead to alerts). Further, the Shape GD is a batch processing algorithm,
thus, the individual nodes can batch their data at coarse time-scales (e.g.
once every NTW) and send the data to the Shape GD.  Finally, it does not matter
even if some batches are lost/missed; recall that the Shape GD is robust to
precisely this type of noise.  
Appendix~A
discusses overheads in more depth but the key takeaway is that
Shape GD has low overheads -- 
each LD can use simple dot products and scalar comparisons to implement PCA and
Random Forests, the total incoming bandwidth to the Shape GD server ranges from
40KBps to 174KBps for phishing and waterhole respectively, and the server only
needs to bin data (into 50 bins) and compute Wasserstein distance (add 10
counters in each bin).





\begin{figure}[t]
\includegraphics[width=0.48\textwidth]{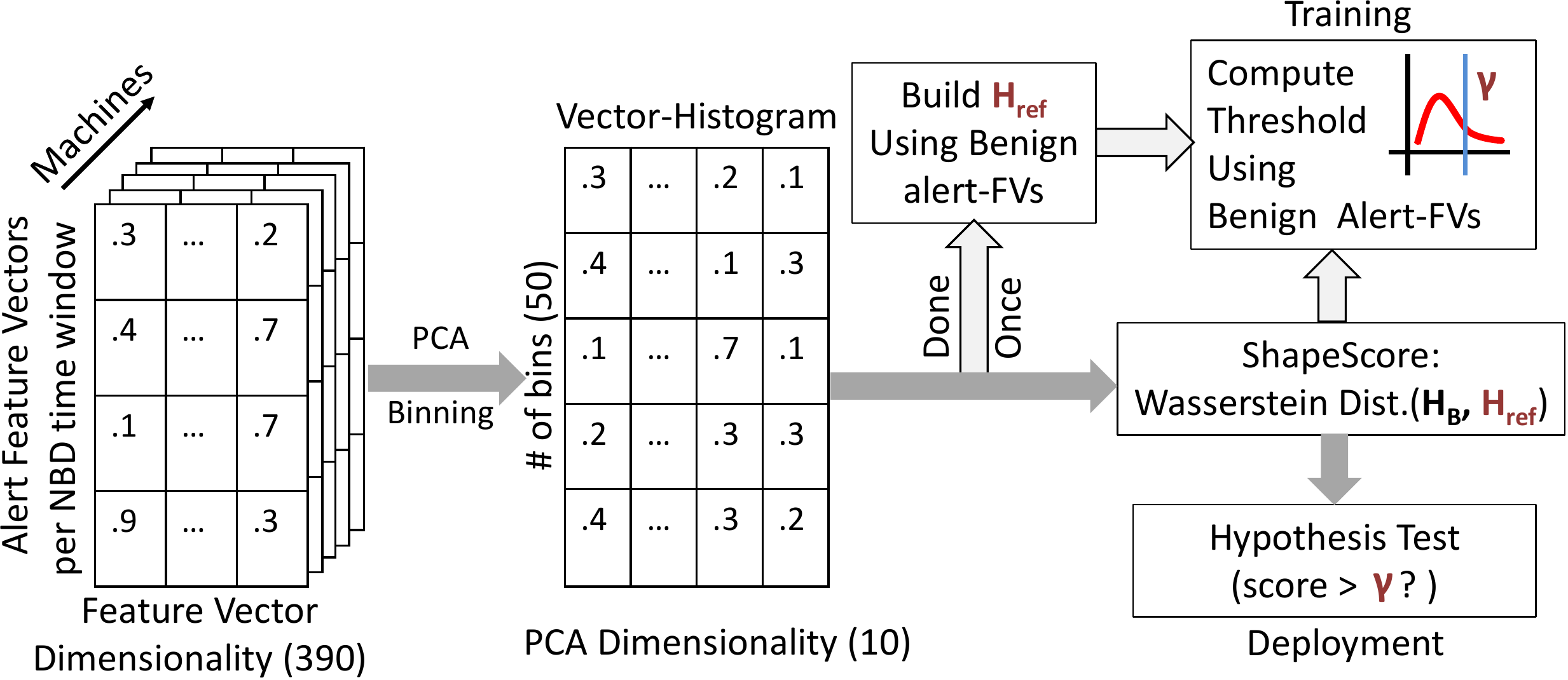}\par
\caption{(Overview) Shape-GD machine learning pipeline.}
\vspace{-0.2in}
\label{fig:shape-gd-pipeline}
\end{figure}

\noindent{\bf Detailed Shape-GD Pipeline.}
As an extension to the description in Section 4, 
Figure~\ref{fig:shape-gd-pipeline} shows the 
detailed machine learning pipeline for extracting 
one neighborhood's shape into a ShapeScore.

\section{How fragile is Count GD to errors in estimating neighborhood size? (Phishing)}
\label{sec:phishing-countgd}

\begin{figure}[tbp]
   \centering
   \includegraphics[width=0.45\textwidth]{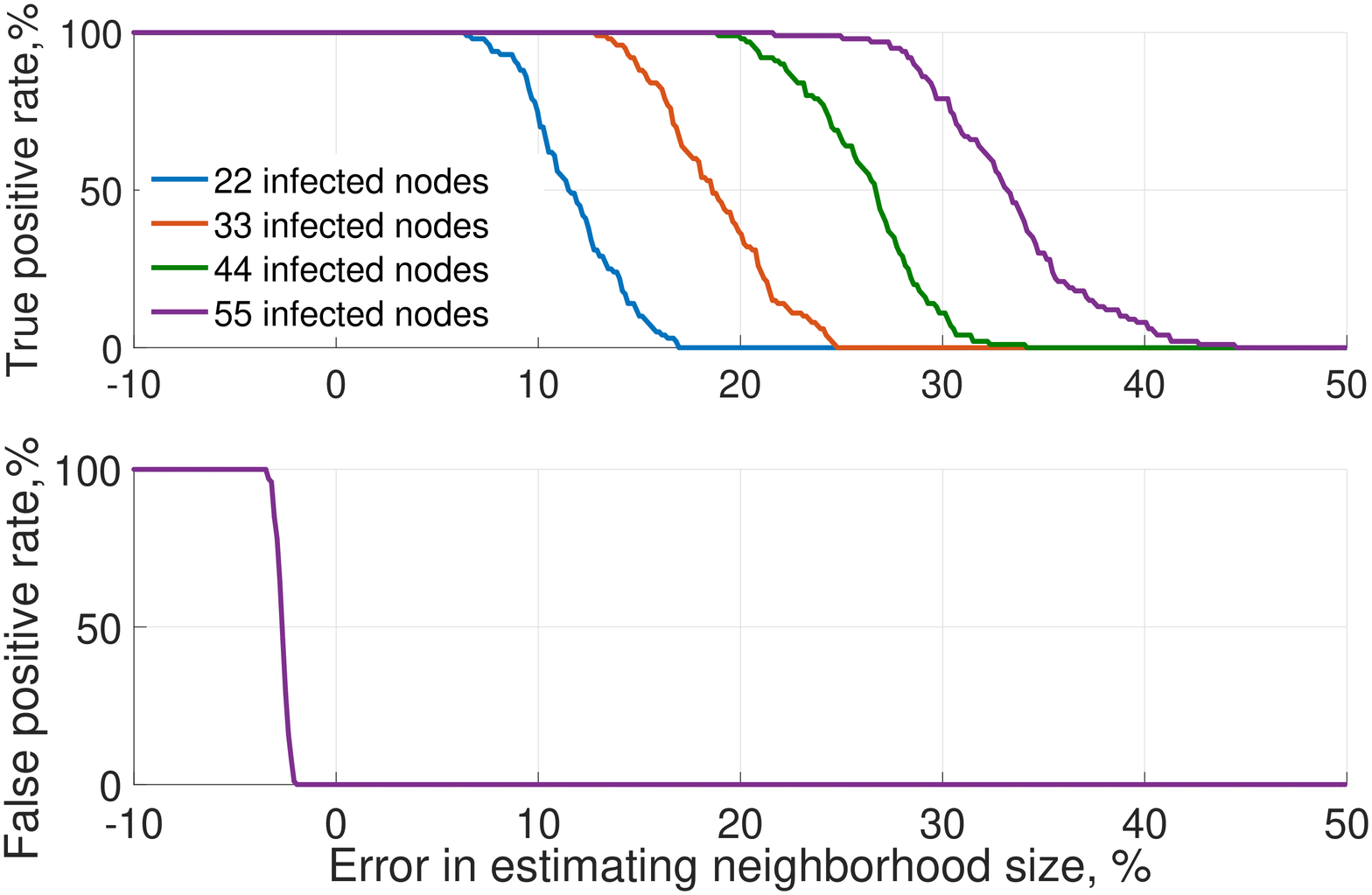}\par
\caption{{\em(Phishing attack) An error in estimating neighborhood size dramatically affects Count GD's performance. 
It can tolerate at most 2\% underestimation errors and 6.3\% overestimation errors to achieve comparable with Shape GD performance.}
\vspace{-0.2in}
\label{fig:windows-noisy_count}}
\end{figure}

Due to space considerations, we have placed the results for Count GD's 
fragility to mis-estimated neighborhood sizes in the appendix in Figure~\ref{fig:windows-noisy_count}.
The key trends are similar to waterhole attack presented in the main paper -- 
even a slight under- or over-estimate of dynamic neighborhoods' sizes
can yield completely inaccurate results.

\end{document}